\documentclass[prb,reprint,twocolumn,superscriptaddress]{revtex4}
\usepackage{graphicx}
\usepackage{amsmath}
\usepackage{amssymb}
\usepackage{hyperref}
\usepackage{pgf}

\newcommand*{\parallelogramm}
{
  \rlap{\rotatebox{-30}{\rule[.05ex]{.4pt}{.77em}}}
  \kern.04em
  \rlap{\kern.36em\raisebox{0.649519052835em}{\rule{.6em}{.4pt}}}
  \rule{.6em}{.4pt}\kern-.04em
  \rotatebox{-30}{\rule[.05ex]{.4pt}{.77em}}}

\begin{document}

\title{Unified tensor network theory for frustrated classical spin models in
two dimensions}
\author{Feng-Feng Song}
\thanks{These authors contributed equally.}
\affiliation{State Key Laboratory of Low Dimensional Quantum Physics and Department of
Physics, Tsinghua University, Beijing 100084, China}
\author{Tong-Yu Lin}
\thanks{These authors contributed equally.}
\affiliation{State Key Laboratory of Low Dimensional Quantum Physics and Department of
Physics, Tsinghua University, Beijing 100084, China}
\author{Guang-Ming Zhang}
\email{gmzhang@tsinghua.edu.cn}
\affiliation{State Key Laboratory of Low Dimensional Quantum Physics and Department of
Physics, Tsinghua University, Beijing 100084, China}
\affiliation{Collaborative Innovation Center of Quantum Matter, Beijing 100084, China}
\affiliation{Frontier Science Center for Quantum Information, Beijing 100084, China}
\date{\today }

\begin{abstract}
Frustration is a ubiquitous phenomenon in many-body physics that influences
the nature of the system in a profound way with exotic emergent behavior.
Despite its long research history, the analytical or numerical
investigations on frustrated spin models remain a formidable challenge due
to their extensive ground state degeneracy. In this work, we propose a
unified tensor network theory to numerically solve the frustrated classical
spin models on various two-dimensional (2D) lattice geometry with high
efficiency. We show that the appropriate encoding of emergent degrees of
freedom in each local tensor is of crucial importance in the construction of
the infinite tensor network representation of the partition function. The
frustrations are thus relieved through the effective interactions between
emergent local degrees of freedom. Then the partition function is written as
a product of a one-dimensional (1D) transfer operator, whose eigen-equation
can be solved by the standard algorithm of matrix product states rigorously,
and various phase transitions can be accurately determined from the
singularities of the entanglement entropy of the 1D quantum correspondence.
We demonstrated the power of our unified theory by numerically solving 2D
fully frustrated XY spin models on the kagome, square and triangular
lattices, giving rise to a variety of thermal phase transitions from
infinite-order Brezinskii-Kosterlitz-Thouless transitions, second-order
transitions, to first-order phase transitions. Our approach holds the
potential application to other types of frustrated classical systems like
Heisenberg spin antiferromagnets.
\end{abstract}

\maketitle

\section{Introduction}

Frustrated spin systems have become an extremely active field of theoretical
and experimental research in the last decades characterized by complex
low-energy physics and fascinating emergent phenomena\cite%
{Lacroix_2011,Ramirez_1994,Moessner_2001}. A system is regarded as
frustrated when conflicting interaction terms are present, featured by the
inability to minimize total energy by concurrently reducing the energy of
each group of interacting degrees of freedom. Frustration underlies
non-trivial behavior across physical systems or more general many-body
systems, as the minimization of local conflicts gives rise to new degrees of
freedom\cite{Diep_2020,Ortiz-Ambriz_2019}.

Classical frustrated spin systems can be understood as simplified quantum
mechanical models which employ classical spins to investigate the behavior
of strongly correlated magnetic systems with competing interactions. The
existence of frustration depends on the lattice geometry and/or the nature
of the interactions\cite{Sadoc_1999}. For example, the anti-ferromagnetic
(AF) Ising model defined by a set of spins of $s=\pm 1$ is frustrated on the
triangular and kagome lattices with massive ground-state degeneracy\cite%
{Wannier_1950,Kano_1953}. However, AF Ising models are not frustrated on the
2D square lattice because the lattice is bipartite and the energy can be
simply minimized by the Neel configuration of alternating spins. Frustration
also depends on the dimension of the spin variables. For the frustrated AF
XY spin systems composed of planar vectors $\vec{s}=(\sin \theta ,\cos
\theta )$, the ground-state configuration is usually highly degenerate with
new symmetries induced from non-collinear patterns. The new degrees of
freedom can give rise to rich and complex phases at finite temperatures,
which have been studied over the past decades on the square\cite%
{Teitel1983,Thijssen1990,Santiago1992,Granato1993,LeeJR1994,LeeSy1994,
Santiago1994, Olsson1995,Cataudella1996,Olsson1997,Boubcheur1998,
Hasenbusch2005,Okumura2011,Nussinov_2014,Lima2019,Song_2022}, the triangular
\cite{Miyashita1984,Shih1984,DHLee1984,DHLee1986,Korshunov1986,Himbergen1986,
Xu1996,LeeSy1998,Luca1998} and the kagome lattices \cite{Harris_1992,
Rzchowski_1997,Cherepanov_2001,Park_2001,Korshunov_2002,Andreanov_2020,Song_2023_2}.

The study of frustrated classical spin systems is important not only for
understanding the emergent behavior of physical systems like spin glasses%
\cite{Villain1977,Binder_1986} but also for general optimization problems
across multiple disciplines\cite{Hartmann_2001}. Considerable efforts have
been made in the investigation of the fundamental properties of frustrated
classical spin systems. Despite decade-long efforts, a generic approach to
dealing with frustrated spin systems with both high accuracy and high
efficiency is still lacking. Well-established methods such as Monte Carlo
simulations, mean-field theories, and renormalization group techniques, have
made significant contributions to the study of the classical frustrated spin
models. However, they have encountered many difficulties such as low
efficiency or limited applications\cite%
{Swendsen_1987,Wolff_1989,Rakala_2017,Andreanov_2020}.

Recent progress in the tensor network methods provides new computational
approaches for studying classical lattice models with strong frustrations%
\cite{Vanderstraeten2018,Vanhecke2021,Song_2022,Colbois_2022,Song_2023_2}.
It is found that the construction of the tensor network of the partition
function is nontrivial for frustrated systems compared to the standard
formulation. For example, the ground state local rules should be encoded in
the local tensors to satisfy the ground state configurations induced by
geometrical frustrations\cite{Vanderstraeten2018}. In the frustrated Ising
models, a linear searching algorithm based on a Hamiltonian tessellation has
been proposed to find the proper transitional invariant unit\cite%
{Vanhecke2021,Colbois_2022}. In the frustrated XY models, the idea of
splitting of $U(1)$ spins and dual transformations have been developed to
overcome the convergence issues\cite{Song_2022,Song_2023_2}. Although these
techniques make a success in specific models, they seem to be very tricky.
Thus, one wonders whether there exists a general framework to treat
frustrated classical spin models.

Here, we generalize the underlying principles of the tensor network
representation to make it applicable to generic frustrated classical spin
systems. When comprising the whole tensor network of the partition function,
the crucial point is that the emergent degrees of freedom induced by
frustrations should be encoded in the local tensors. In this way, the
massive degeneracy is characterized by emergent dual variables such as
height variables in the AF Ising model on the triangular lattice\cite%
{Blote_1982,Chalker_2014} and chiralities in frustrated XY models\cite%
{Korshunov_2002,Song_2023_2}. The emergent variables capture the freedom of
a group of interacting spins under the constraint of frustrations. In the
sense of coarse-graining, the local tensors carry the effective interactions
between emergent local degrees of freedom. The local tensors usually sit on
the dual sites of the original lattice which can be constructed from dual
transformations. It is worth noting that the dual transformations should be
imposed on the whole cluster of a number of spins in correspondence with the
emergent dual variables.

We demonstrate the power of the generalized theory of tensor network
representation by applying it to fully frustrated XY models on the kagome,
triangular, and square lattices. First of all, we can express the infinite
2D tensor network as a product of 1D transfer matrix operators, which can be
contracted efficiently by recently developed tensor network algorithms under
optimal variational principles\cite%
{Stauber_2018,Vanderstraeten_2019,Nietner_2020}. Then, from the singularity
of the entanglement entropy of the 1D quantum transfer operator, various
phase transitions can be determined with great accuracy according to the
same criterion\cite{Haegeman_2017}. Finally we find that a broad array of
emergent physics has been treated including various types of phase
transitions from first-order, second-order to the
Berezinskii-Kosterlitz-Thouless (BKT) phase transitions. The complex phase
structures of the frustrated XY systems are revisited and clarified with new
tensor network solutions. The present approach holds the potential
application to next-nearest-neighbor frustrated spin systems and other types
of classical spins like Heisenberg antiferromagnet.

The rest of the paper is organized as follows. In Sec. II, we introduce the
theory of tensor network representations for classical frustrated spin
models with two concrete examples. After constructing the tensor networks of
Ising spin antiferromagnets on the kagome and triangular lattices, we
perform the numerical calculation of the residual entropy of the frustrated
Ising models, which are comparable to the exact results. In Sec. III, we
apply the unified theory to the fully frustrated XY spin models on the
kagome, square, and triangular lattices, and present the numerical results
for the determination of the finite temperature phase diagram of frustrated
XY systems, especially the AF triangular XY model and the modified square XY
model. Finally in Sec. IV, we discuss the future generalizations of the
method and give our conclusions. In the Appendix, we outline the detailed
tensor network methods for numerical calculations.

\section{Tensor network representations of 2D statistical models}

\subsection{Emergent degrees of freedom}

Tensor networks have proven to be a very potent tool in the study of
strongly correlated quantum models as well as classical statistical
mechanics. To implement this powerful method, the first step is to convert
the partition function of a classical lattice model with local interactions
into a tensor network representation.

The standard construction of the tensor network is conducted by putting a
matrix on each bond of the original lattice accounting for the Boltzmann
weight of the nearest-neighboring interactions\cite{Zhao_2010}. For a
generic spin model with nearest-neighbor interactions
\begin{equation}
H=\sum_{\langle i,j\rangle}h(s_i,s_j),
\end{equation}
the partition function can be decomposed into a tensor network as a product
of local Boltzmann weights,
\begin{equation}
Z=\sum_{\{s_i\}}\mathrm{e}^{-\beta H(\{s_i\})}=\sum_{\{s_i\}}\prod_{\langle
i,j\rangle}W(s_i,s_j),
\end{equation}
where $\langle i,j\rangle$ refers to the nearest neighbors, $s_i$ are the
spin variables, and the interaction matrices are given by
\begin{equation}
W(s_i,s_j)=\mathrm{e}^{-\beta h(s_i,s_j)},
\end{equation}
whose row and column indices are the spin variables shown in Fig.~\ref%
{fig:stand_rep}. The $\delta$ tensors on the lattice vertexes ensure all
indices of $W$ take the same value at the joint point.

\begin{figure}[tbp]
\centering
\includegraphics[width=\linewidth]{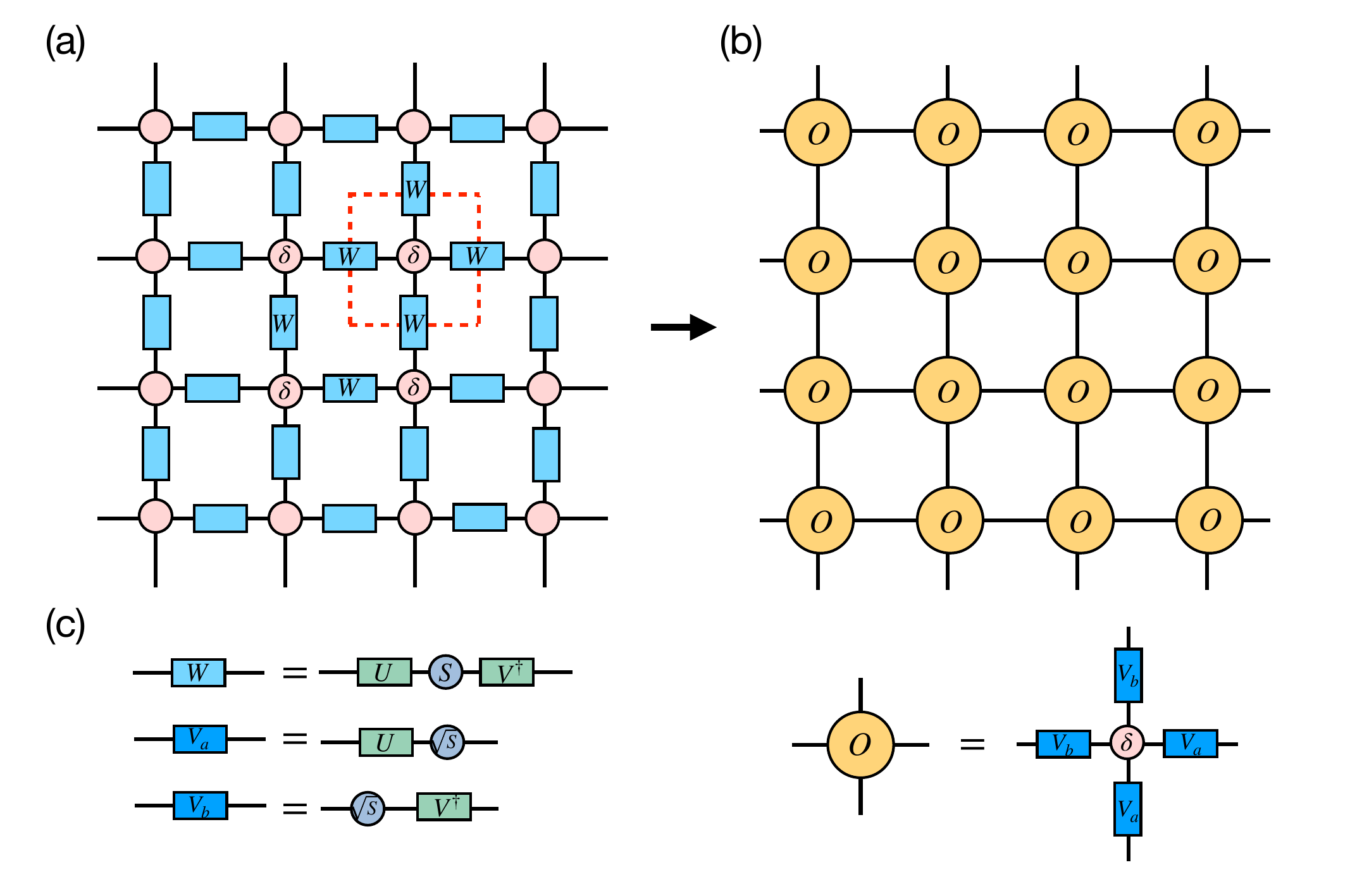}
\caption{ The standard construction of the tensor network. (a) The $W$
matrix represents the Boltzmann weight on each link, and the $\protect\delta$
tensor on each site represents the sharing of the same spin between
neighboring $W$ matrices. (b) The tensor network representation of the
partition function composed of uniform local tensors. (c) The local tensor $O
$ is built by the singular value decomposition (SVD) on each $W$ matrix and
the grouping of the $V$ matrices connecting to the $\protect\delta$ tensors.
}
\label{fig:stand_rep}
\end{figure}

Furthermore, we perform the Schmidt decomposition on the symmetric matrix $W$
\begin{equation}
W(s_{i},s_{j})=(U\sqrt{S})(\sqrt{S}V^{\dagger
})=V_{a}(s_{i},s_{k})V_{b}(s_{k},s_{j}),
\end{equation}%
and the partition function can be cast into the uniform tensor network
representation as shown in Fig.~\ref{fig:stand_rep}
\begin{equation}
Z=\mathrm{tTr}\prod_{i}O_{s_{1},s_{2}}^{s_{3},s_{4}}(i)
\end{equation}%
by grouping all V matrices that connect to the $\delta $ tensors
\begin{equation}
O_{s_{1},s_{2}}^{s_{3},s_{4}}=%
\sum_{s_{k}}V_{b}(s_{1},s_{k})V_{b}(s_{2},s_{k})V_{a}(s_{k},s_{3})V_{a}(s_{k},s_{4}).
\end{equation}

The standard representation has been successfully applied to many lattice
statistical models without frustration\cite%
{Levin_2007,Zhao_2010,Yu_2014,Haegeman_2017,Vanderstraeten2019_2}. However,
it cannot be implemented directly in the frustrated spin models, where the
tensor network contraction algorithms fail to converge. It was found that
the proper encoding of the ground state local rules in local tensors was
crucial for the contraction to converge. To fulfill the physics of the
ground state manifold, a linear algorithm was proposed to search for the
optimal Hamiltonian tessellation for Ising antiferromagnets\cite%
{Vanhecke2021,Colbois_2022}. The key point is that the energy of all local
ground state configurations should be simultaneously minimized under the
splitting of the global Hamiltonian into local groups of interactions. And
the local tensors are constructed as translational units coinciding with the
local clusters of the tessellation.

In order to extend tensor network approaches to generic frustrated classical
spin models, we should understand the ground state local rules from a more
fundamental perspective of emergent degrees of freedom. In frustrated
systems, new degrees of freedom often emerge as a result of the minimization
of local conflicts. The ground state of frustrated spin systems is highly
degenerate because a number of spins can behave as free spins. Such freedom
can therefore be represented by a set of emergent variables describing the
effective interactions induced by frustrations. For some models, the
emergent variables can be derived directly like height variables in the AF
Ising triangular model\cite{Blote_1982,Chalker_2014} and chiralities in
frustrated XY models\cite{Korshunov_2002,Song_2023_2}. For the spin models
with more complicated interactions, the emergent variables may not be
explicitly expressed but they can still be characterized by local tensors
composed of a cluster of local interactions\cite{Vanhecke2021,Colbois_2022}.
This idea generalizes tensor network approaches readily to classical
frustrated systems of both discrete and continuous spins.

\begin{figure*}[tbp]
\centering
\includegraphics[width=\linewidth]{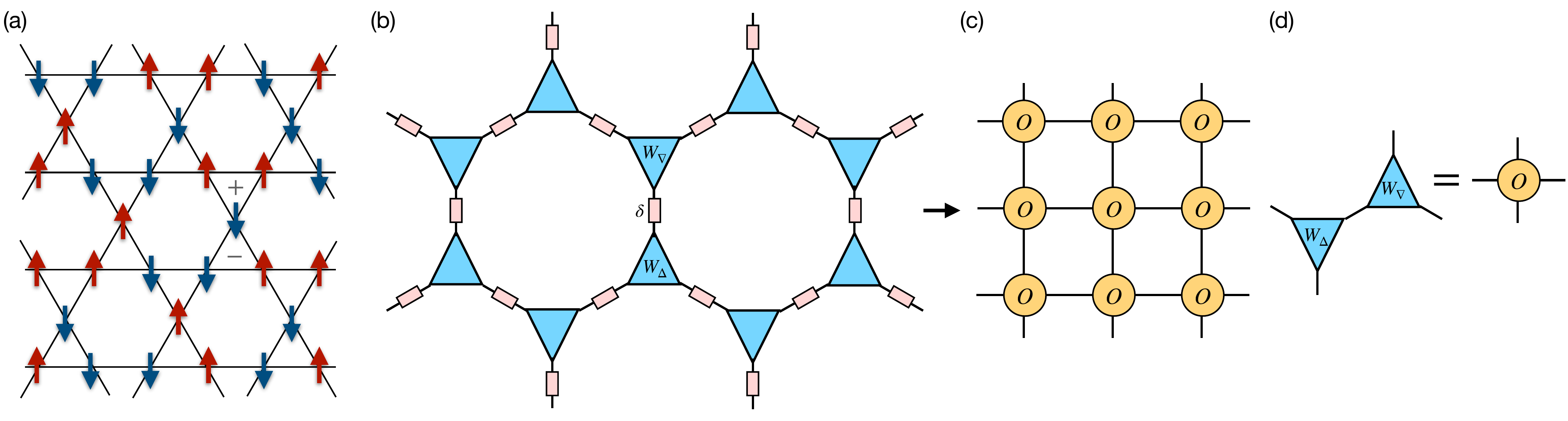}
\caption{ Tensor network representation of the AF Ising model on a kagome
lattice. (a) One of the ground state configurations on the kagome lattice
with $Q=\pm 1$ charges on each triangle. (b) Putting the $W_{\Delta}$ ($W_{%
\protect\nabla}$) tensors on the centers of the upward (downward) triangles
to represent the self-energy of the charge variables, where the $\protect%
\delta$ tensors between the nearest neighbor triangles can be translated
into the connections of tensor legs directly. (c) The tensor network
representation of the partition function composed of uniform local $O$
tensors. (d) The construction of $O$ tensor by contracting neighboring $%
W_{\Delta}$ and $W_{\protect\nabla}$ tensors.}
\label{fig:kgm_Ising}
\end{figure*}

Before discussing the tensor network construction of the frustrated spin
model, we give some examples of emergent degrees of freedom by revisiting
the exactly solvable frustrated models. One of the simplest frustrated spin
models is the AF Ising model on the kagome lattice
\begin{equation}
H=J\sum_{\langle i,j\rangle }\sigma _{i}\sigma _{j},  \label{eq:kgm_Ising}
\end{equation}%
where $J>0$ denotes the AF interactions between nearest-neighbor spins $%
s_{i}=\pm 1$ as displayed in Fig.~\ref{fig:kgm_Ising} (a).

The kagome AF Ising model is disordered at all temperatures with an
extensive ground state degeneracy characterized by a finite residual entropy%
\cite{Kano_1953}. To minimize the energy of each triangular plaquette, three
spins should obey the ground state local rule of \textquotedblleft two up
one down, one down two up\textquotedblright\ as shown in Fig.~\ref%
{fig:kgm_Ising} (a).

Besides directly focusing on the local spin configurations, the physics of
the model can be understood from the emergent degrees of freedom on the
triangle centers. A set of charge variables can be defined at each triangle
\begin{equation}
Q_{u}=\sum_{i\in \Delta }s_{i},\quad Q_{d}=-\sum_{i\in \nabla }s_{i},
\label{eq:kgm_charge}
\end{equation}%
where $\Delta $ and $\nabla $ denote the upward and downward triangles. The
Hamiltonian can then be expressed as
\begin{equation}
H=\frac{J}{2}\sum_{p\in \Delta (\nabla )}(Q_{p}^{2}-3)
\end{equation}
in terms of the topological charges $Q_{p}$.

Although there seems to be no explicit interaction between charges in the
Hamiltonian, the variables $Q_{p}$ are not independent because the shared
spin between the neighboring triangles should be the same. The constraints
between neighboring charges can be naturally represented by a link between
local tensors as a Kronecker delta tensor in the language of tensor
networks. In this way, the interactions between Ising spins are transformed
into a charge model including the self-energy of the charges and the
effective interactions between these charges. The charge variables can take
four values $Q=\pm 1,\pm 3$ at finite temperatures. In the zero temperature
limit, the charges of $Q=\pm 3$ are energetically suppressed. The
\textquotedblleft two up one down, one up two down\textquotedblright\ rule
corresponds to charge variables $Q=\pm 1$ allowed by the ground state
manifold.

\begin{figure*}[t]
\centering
\includegraphics[width=\linewidth]{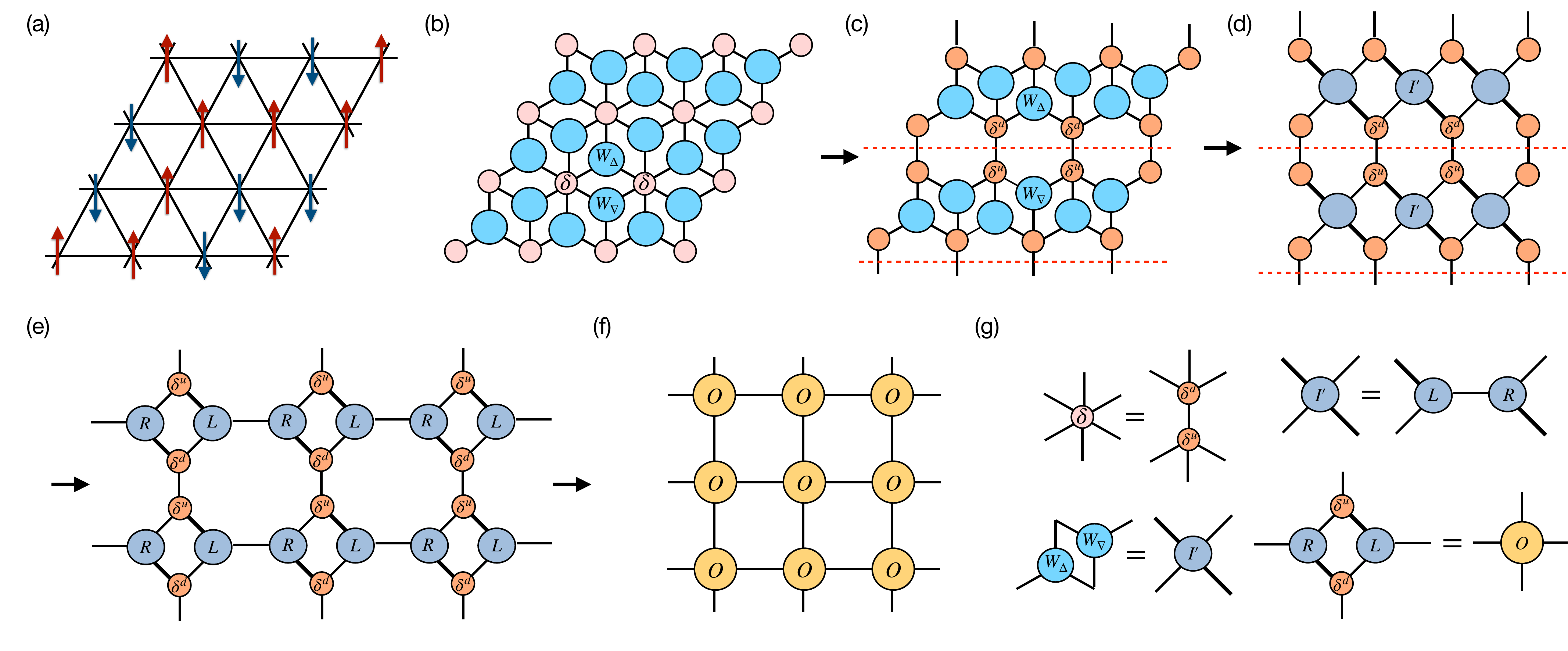}
\caption{ Tensor network representations of the Ising anti-ferromagnet on a
triangular lattice. (a) One of the massive degenerate ground state
configurations. (b) The $W_{\protect\nabla}$ and $W_{\Delta}$ tensors are
defined on the center of the triangles. The pink $\protect\delta$ tensor
represents a six-legged Kronecker delta tensor which connects the $W_{%
\protect\nabla}$ and $W_{\Delta}$ tensors surrounding it. (c)-(d) The
construction of row-to-row transfer matrix by splitting the six-legged $%
\protect\delta$ tensors vertically and regrouping the index of a pair of
neighboring $W_{\protect\nabla}$ and $W_{\Delta}$ tensor into an $I^{\prime }
$ tensor. (e)-(f) The construction of the local uniform tensor O by
splitting $I^{\prime }$ horizontally and grouping with $\protect\delta^u$
and $\protect\delta^d$ tensors. (g) The details of the operations on local
tensors during the construction procedure.}
\label{fig:tri_Ising}
\end{figure*}

The emergent charge variables can also be applied to the triangular lattice
in the same spirit as the case of the kagome lattice. The triangular AF
Ising model in Fig.~\ref{fig:tri_Ising} (a) can be transformed into
\begin{equation}
H=\frac{J}{2}\sum_{\langle i,j\rangle \in p}s_{i}s_{j}=\frac{J}{4}%
\sum_{p}\left( Q_{p}^{2}-3\right) ,
\end{equation}%
where the only difference is that each nearest-neighbor triangles share two
same spins. The charges variables help us to understand why the tiling of $%
p\in \Delta (\nabla )$ is crucial for the triangular lattices\cite%
{Vanhecke2021}. The reason is that the tessellation of only one type of
triangle fails to characterize the interactions between the emergent charge
variables.

\subsection{General principle for tensor network construction}

Now we can build up a general principle for the tensor network
representation of frustrated spin models. The key point is that the emergent
degrees of freedom should be encoded in each local tensor in the
construction of the infinite tensor network for the partition function.
Since the emergent degrees of freedom is universal in frustrated systems,
the generic approach can be applied to classical frustrated systems of both
discrete and continuous symmetries. Moreover, the finite-temperature
properties can also be probed when the interactions among emergent degrees
of freedom are faithfully captured.

In practice, it is not necessary to write down the explicit model of the
interactions between emergent variables. The effective interactions are
implicit in the connections between local tensors. Each local tensor
constituting the Boltzmann weight should carry the emergent degrees of
freedom corresponding to a unit cluster of spins. From this perspective, the
breakdown of standard construction in the triangular Ising model\cite%
{Vanhecke2021} can be understood: the emergent degrees of freedom located on
the downward triangles are lost in the infinite tensor network contraction.

We summarize the general procedure to construct the tensor network
representation of the frustrated spin models as follows:

i). Identify the emergent degree of freedom, usually located on the dual
site, and the corresponding geometry cluster composed of classical spins.

ii). Reformulate the partition function into the form of
\begin{equation}
Z=\sum_{\{s\}}\prod_{c}W_c(c)\prod_{\langle c,c^{\prime
}\rangle}W_l(c,c^{\prime })\delta_{c,c^{\prime }}
\end{equation}
where $c$ enumerates all the clusters, $W_c$ and $W_l$ correspond to the
Boltzmann weight of all the spin configurations $\{s\}$ within a cluster and
between neighboring clusters, and $\delta$ tensors ensure the shared spins
between different clusters be the same. For continuous spins, the $W$
tensors should be transformed onto a discrete basis via the Fourier
transformation.

iii). Split and regroup the $W$ tensors to build regular local tensors
constituting an infinite uniform tensor network representation of the
partition function.

\subsection{Kagome and triangular AF Ising models as two examples}

The general principle can be applied directly to classical frustrated models
with discrete symmetries. The tensor network representation of the kagome AF
Ising model \eqref{eq:kgm_Ising} can be built simply based on the emergent
charge variables defined in \eqref{eq:kgm_charge}. As displayed in Fig.~\ref%
{fig:kgm_Ising} (b), we first split the global Boltzmann weight into local
Boltzmann weights on each triangle. Then the partition function of the AF
Ising model can be written as
\begin{equation}
Z=\sum_{\{s_{i}\}}\prod_{p}W_{p}(s_{1},s_{2},s_{3}),
\end{equation}%
where the Boltzmann weight on each upward and downward triangle is expressed
by a three-legged $W$ tensor
\begin{equation}
W_{p}(s_{1},s_{2},s_{3})=\mathrm{e}^{-\beta
J(s_{1}s_{2}+s_{2}s_{3}+s_{3}s_{1})}.  \label{eq:kgm_Ising_W}
\end{equation}%
The constraint of sharing the same spin between a pair of neighboring $W$
tensors is imposed by the Kronecker delta tensor.

Then the transitional invariant local tensor $O$ is achieved by combining a
pair of upward and downward triangles
\begin{equation}
O_{s_1,s_2}^{s_3,s_4}=\sum_{s_5}
W_{\Delta}(s_1,s_2,s_5)W_{\nabla}(s_5,s_3,s_4)
\end{equation}
as displayed in Fig.~\ref{fig:kgm_Ising} (d), and the uniform tensor network
representation of the partition function in Fig.~\ref{fig:kgm_Ising} (c) is
given by
\begin{equation}
Z=\mathrm{tTr}\prod_i O_{s_1,s_2}^{s_3,s_4}(i)
\end{equation}
where ``tTr means the tensor contraction over all auxiliary links and $i$
denotes the sites of the transitional invariant unit.

The above tensor network can be contracted efficiently using standard
algorithms for infinite systems with extremely high accuracy\cite%
{Haegeman_2017,Stauber_2018,Vanderstraeten_2019}. In the zero temperature
limit, the tensor $W$ can be reduced to the same tensor obtained in the Ref.%
\cite{Vanhecke2021}, yielding a residual entropy of $S_{0}\approx 0.501833$,
consistent with the exact result\cite{Kano_1953}.

For the triangular AF Ising model displayed in Fig.~\ref{fig:tri_Ising} (a),
the tensor network representation can be constructed in a similar way. The
only difference is that each spin is shared by six surrounding triangles. As
shown in Fig.~\ref{fig:tri_Ising} (b), the constraint between the triangular
plaquettes is realized through the six-legged delta tensors
\begin{equation}  \label{eq:kroneck_delta}
\delta _{s_{1},s_{2},s_{3},s_{4},s_{5},s_{6}}=
\begin{cases}
1, & s_{1}=s_{2}=s_{3}=s_{4}=s_{5}=s_{6} \\
0, & \text{otherwise}%
\end{cases}%
\end{equation}
and the tensor $W$ is defined in the same way as the kagome AF Ising model
Eq.~\eqref{eq:kgm_Ising_W}.

To construct a row-to-row transfer matrix, we split the six-legged delta
tensors vertically as two four-legged delta tensors
\begin{equation}
\delta _{s_{1},s_{2},s_{3},s_{4},s_{5},s_{6}}=\sum_{s_{7}=\pm 1}\delta
_{s_{1},s_{2},s_{3},s_{7}}^{u}\delta _{s_{7},s_{4},s_{5},s_{6}}^{d}
\end{equation}%
as shown in Fig.~\ref{fig:tri_Ising} (c). Then a pair of $W_{\Delta }$ and $%
W_{\nabla }$ are grouped into a tensor $I^{\prime }$ as shown in Fig.~\ref%
{fig:tri_Ising} (d). The tensor $I^{\prime }$ can be further split
horizontally as displayed in Fig.~\ref{fig:tri_Ising} (e)
\begin{equation}
I^{\prime }=LR
\end{equation}%
by a singular-value decomposition
\begin{equation}
I^{\prime }=USV^{\dagger},
\end{equation}%
where $U$ and $V^{\dagger}$ are three-legged unitary tensors, $S$ is a
semi-positive diagonal matrix and
\begin{equation}
L=U\sqrt{S},\quad R=\sqrt{S}V^{\dagger}.
\end{equation}

Finally, the regular local tensor $O$ is obtained by grouping $\delta^u$, $%
\delta^d$, and a pair of $L$ and $R$ tensors. The details are depicted in
Fig.~\ref{fig:tri_Ising} (g). This gives a uniform tensor-network
representation of the partition function
\begin{equation}
Z=\mathrm{tTr}\prod_i O_{s_1,s_2}^{s_3,s_4}(i)
\end{equation}
as displayed in Fig.~\ref{fig:tri_Ising} (f). Although the local tensor $O$
is slightly different from the one constructed by the method of Hamiltonian
tessellation\cite{Vanhecke2021}, the tensor network is well defined and can
be readily generalized to frustrated systems with continuous symmetries
discussed in the following parts.

\begin{figure}[t]
\centering
\includegraphics[width=\linewidth]{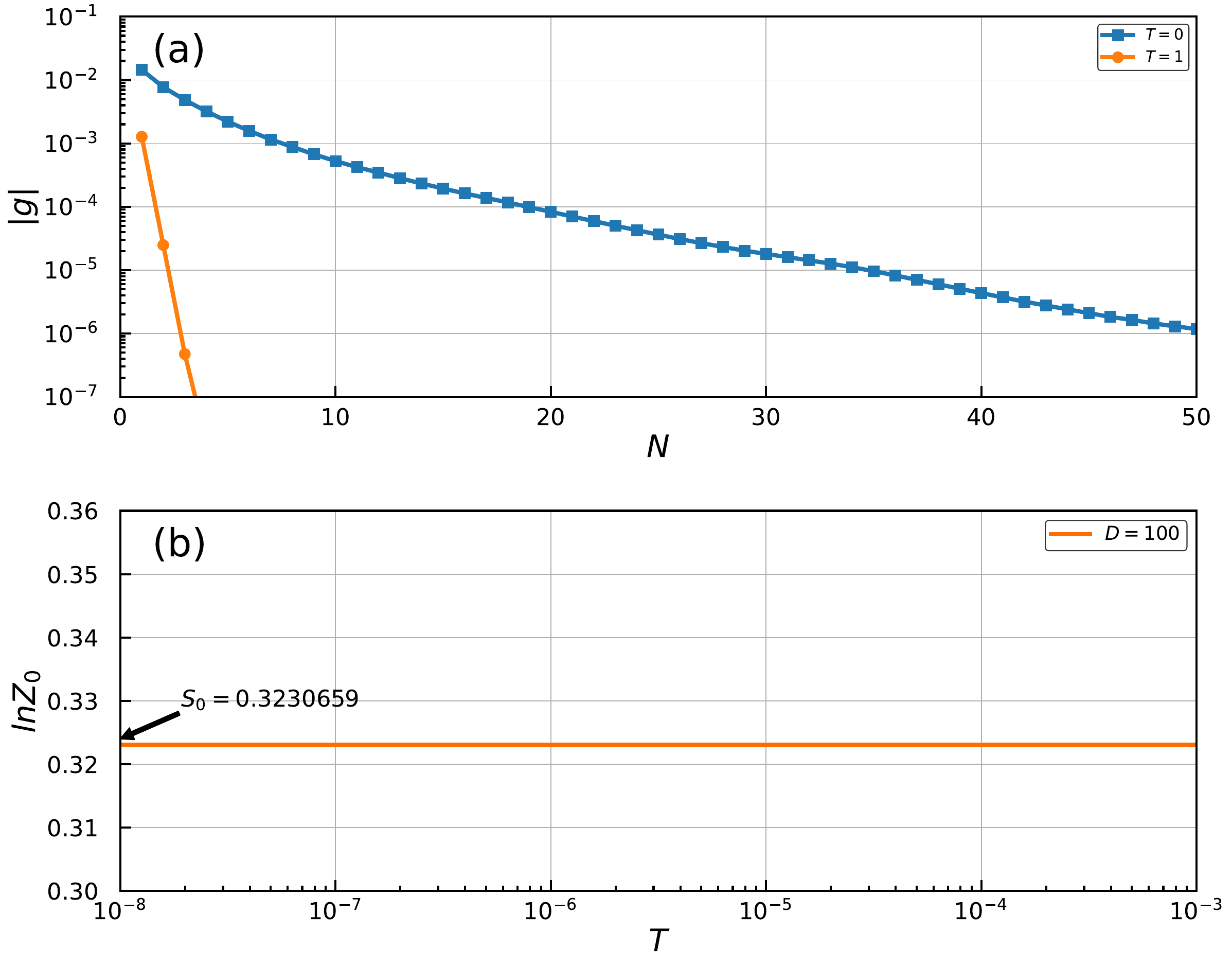}
\caption{ Numerical results of the Ising anti-ferromagnet on the triangular
lattice, the bond dimension of uniform MPS is $D=100$. (a) Convergence of
the VUMPS algorithm at $T=1$ and $T=0$. $|g|$ is the convergence measure in
the VUMPS algorithm and $N$ is the number of iteration steps. (b) $\ln Z_0$
as a function of temperature. The residual entropy per site is $%
S_0(D=100)=0.3230659$, which is the same as the exact result to seven
decimal places. }
\label{fig:tri_Ising_result}
\end{figure}

As shown in Fig.~\ref{fig:tri_Ising_result} (a), standard contraction
algorithms\cite{Stauber_2018,Fishman_2018,Vanderstraeten_2019} display a
nice convergence at both zero temperature and finite temperatures. The
numerical calculation of the expectation value of the magnetization
\begin{equation}
m=\langle s_i\rangle=\frac{1}{N}\sum_i s_i
\end{equation}
is found to be zero under all temperatures, indicating the absence of the
long-range order (LRO). Moreover, the ground state residual entropy is
calculated as displayed in Fig.~\ref{fig:tri_Ising_result} (b)
\begin{equation}
S_0=\frac{1}{N}\ln Z_0\approx 0.323065,
\end{equation}
in good agreement with the exact result\cite{Wannier_1950}.

\section{Tensor network theory for 2D fully frustrated XY spin models}

\subsection{Duality transformation and split of $U(1)$ spins}

In this section, we demonstrate the power of the generic idea of emergent
degrees of freedom by the implementations in the frustrated model with a
continuous $U(1)$ symmetry. The frustrated XY models, to some extent, are
``less frustrated" than the Ising ones. The XY spins have more freedom to
rotate on the plane to minimize local conflict interactions, but the Ising
spins are constrained to only two orientations. That is why there exists
quasi-LRO in the frustrated XY spin models at low temperatures, while the
frustrated Ising models are usually disordered even at zero temperature.
Despite a long history of investigations\cite%
{Teitel1983,Thijssen1990,Santiago1992,Granato1993,LeeJR1994,LeeSy1994,
Santiago1994, Olsson1995,Cataudella1996,Olsson1997,Boubcheur1998,
Hasenbusch2005,Okumura2011,Nussinov_2014,Lima2019,Song_2022,Miyashita1984,
Shih1984,DHLee1984,DHLee1986,Korshunov1986,Himbergen1986,Xu1996,LeeSy1998,
Luca1998,Harris_1992,Rzchowski_1997,Cherepanov_2001,Park_2001,Korshunov_2002,
Andreanov_2020,Song_2023_2}, many properties of the frustrated XY spin systems
are still not well understood.

In both frustrated and non-frustrated XY models, a widely accepted and
established analytical tool is the 2D Coulomb gas representation\cite%
{J.M.Kosterlitz_1973,J.M.Kosterlitz_1974,Minnhagen_1987}. However, the form
of Coulomb gas formulation is obtained through an approximate approach\cite%
{J.M.Kosterlitz_1973,J.M.Kosterlitz_1974} and it is hard to directly
represent the charge variables by original phase variables\cite%
{Vallat_1994,Nussinov_2014}. Instead, we can comprehend the topological
charge, located on the dual sites, as a coarse-grained degree of freedom
formed by a cluster of phase variables located on the original plaquette.
This understanding serves as a fundamental perspective for constructing the
tensor network of the frustrated XY spin models.

Our tensor network approach provides a universal tool to deal with
frustrated systems on various lattice geometries. We can reformulate the
partition function into a general form of in the same way as the Ising case
\begin{equation}
Z=\prod_{i}\int\frac{d\theta_i}{2\pi}\prod_{p}W_p(\{\theta_p\})
\label{eq:fxy_g}
\end{equation}
where $p$ denotes the plaquette of the lattice and $W_p$ corresponds to the
Boltzmann weight of the elementary cluster. However, different from the
Ising case studied in the Ref.\cite{Vanhecke2021}, one may encounter two
technical issues when constructing a tensor network based on (\ref{eq:fxy_g}%
). First, the indices of local tensors are continuous spin variables, which
is hard to treat in the framework of tensor networks. So the Fourier
transformation is necessary to bring the local tensors onto a discrete
basis. Second, the Kronecker delta functions describing the constraints of
the sharing spins are changed to the Dirac delta functions. For the Ising
spin cases, the shared spins are split and connected directly by the
Kronecker delta functions. Such a strategy cannot be simply extended for the
case of continuous spins because the loops of the Dirac delta functions are
not well defined. This problem can be overcome by introducing an auxiliary
spin connecting to the shared spins between different clusters.

\subsection{AF XY spin model on a kagome lattice}

\begin{figure*}[tbp]
\centering
\includegraphics[width=\linewidth]{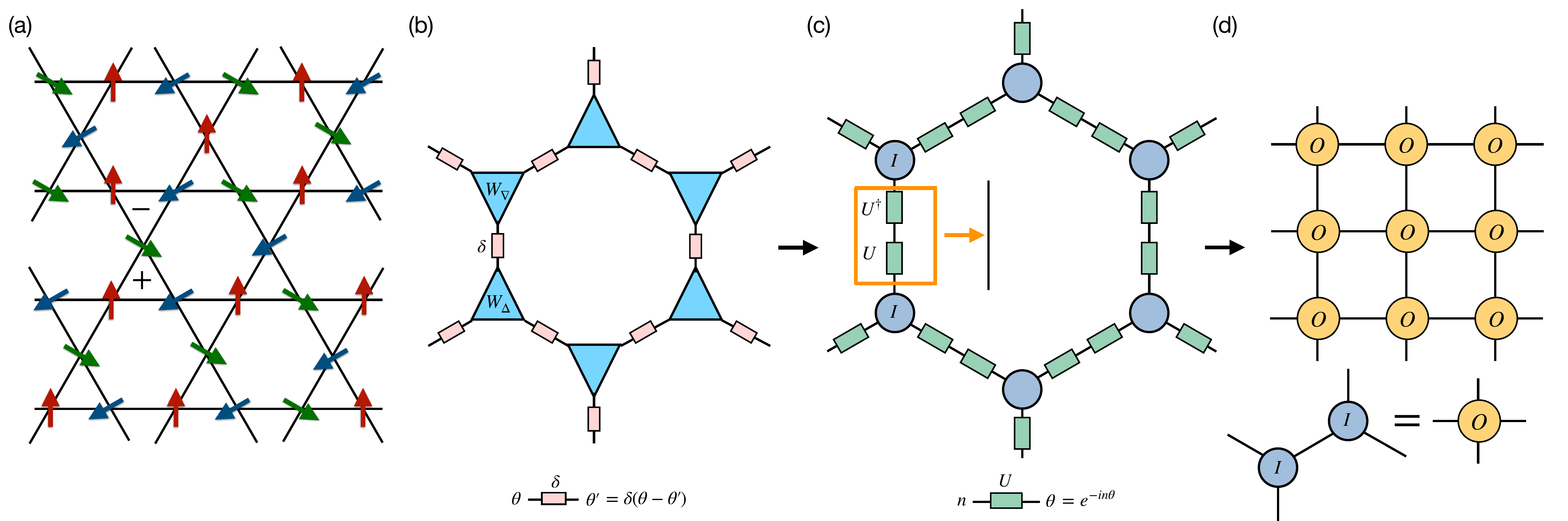}
\caption{ Tensor network representation of the fully frustrated XY model on
the kagome lattice. (a) One of the ground state configurations. The positive
and minus signs denote the chiralities on the triangular plaquettes. (b) The
tensor network with continuous indices. The $W_{\Delta}$ and $W_{\protect%
\nabla}$ tensors represent the Boltzmann weight on up and down-type
triangles. The $\protect\delta$ matrix represents the Dirac delta function.
(c) The construction of tensor network with discrete indices by making
Fourier transformation on each triangle plaquette and integrating out $\{%
\protect\theta\}$ variables. (d) The tensor network representations composed
of local uniform tensor $O$, where the $O$ tensor is built by combining two
neighboring $I$ tensors}
\label{fig:kgm_XY}
\end{figure*}

To describe the Josephson junction array under a uniform external magnetic
field\cite{Teitel1983,Park_2001}, the frustrated XY model on a kagome
lattice in Fig.~\ref{fig:kgm_XY} (a) is defined by the Hamiltonian
\begin{equation}
H=-J\sum_{\langle i,j\rangle}\cos(\theta_i-\theta_j-A_{ij})
\end{equation}
where $J > 0$ is the coupling strength, $i$ and $j$ are the lattice sites,
and the summation is over all pairs of the nearest neighbors. The
frustration in this model is induced by the gauge field defined on the
lattice bond satisfying $A_{ij}=-A_{ji}$. The case of full frustration
corresponds to one-half flux quantum per plaquette,
\begin{equation}
f=\frac{1}{2\pi}\sum_{\langle i,j\rangle\in\Delta}A_{ij}=\frac{1}{2},
\end{equation}
where the sum is taken around the perimeter of a plaquette. We can choose
the fixed gauge condition of $A_{ij}=\pm\pi$ on each bond of the triangular
plaquettes, and the model is transformed into an AF XY model on the kagome
lattice
\begin{equation}
H=J\sum_{i,j}\cos(\theta_i-\theta_j).
\end{equation}

The ground state of this model can be obtained by simultaneously minimizing
the energy on each elementary triangle. As shown in Fig.~\ref{fig:kgm_XY}%
(a), the phase difference between each pair of neighboring spins should be $%
\pm 2\pi /3$. which gives rise to the emergent degrees of freedom of
chiralities $\tau =\pm 1$, corresponding to the anti-clockwise and clockwise
rotation of the spins around the plaquette. The ground state of the AF XY
model on a kagome lattice has a massive accidental degeneracy described by
the fluctuations of the chiralities.

To capture the emergent degrees of freedom induced by frustrations in the
construction of the tensor network, we divide the Hamiltonian into local
terms on each triangle:
\begin{equation}
H=\sum_p H_p,
\end{equation}
where $H_p$ includes all the interactions within an elementary triangle
\begin{equation}
H_p=J\sum_{\langle i,j\rangle \in p}\cos(\theta_i-\theta_j).
\end{equation}
The partition function can now be written as
\begin{equation}
Z=\prod_{i}\int\frac{d\theta_i}{2\pi}\prod_{p}W_p,
\end{equation}
where $W_p=\mathrm{e}^{-\beta H_p}$ is a three-legged tensor with continuous
$U(1)$ indices and the constraint of sharing the same spin at the corners is
realized by the Dirac delta function $\delta(\theta_i-\theta_i^{\prime })$,
as shown in Fig.~\ref{fig:kgm_XY} (b).

To transform the local tensors onto a discrete basis, we employ the duality
transformation to the whole upward triangles
\begin{equation}
I_{n_1,n_2,n_3}=\prod_{i=1}^{3}\int\frac{d\theta_i}{2\pi}W_{\Delta}(%
\theta_1,\theta_2,\theta_3)U_{n_1}(\theta_1)U_{n_2}(\theta_2)U_{n_3}(%
\theta_3),  \notag
\end{equation}
and the downward triangles
\begin{equation}
I^{\prime }_{n_1,n_2,n_3}=\prod_{i=1}^{3}\int\frac{d\theta_i}{2\pi}%
W_{\nabla}(\theta_1,\theta_2,\theta_3)U_{n_1}^{\dagger}(\theta_1)U_{n_2}^{%
\dagger}(\theta_2)U_{n_3}^{\dagger}(\theta_3),  \notag
\end{equation}
where
\begin{equation}
U_n(\theta)=\mathrm{e}^{-in\theta}  \label{eq:Fourier}
\end{equation}
are the basis of the Fourier transformation. Since $W_p$ is unchanged under
the spin reflection of $\theta\to -\theta$, we have $I_{n_1,n_2,n_3}=I^{%
\prime }_{n_1,n_2,n_3}$ as displayed in Fig.~\ref{fig:kgm_XY} (c).
Meanwhile, the duality transformation on the Dirac delta function gives the
Kronecker delta function
\begin{equation}
\int\frac{d\theta}{2\pi}U_{n_1}^{\dagger}(\theta)U_{n_2}(\theta)=%
\delta_{n_1,n_2}.
\end{equation}

Finally, the translation-invariant local tensor $O$ is achieved by combining
a pair of $I$ tensors and we arrive at the the uniform tensor network
representation of the partition function
\begin{equation}
Z=\mathrm{tTr}\prod_i O_{s_1,s_2}^{s_3,s_4}(i)
\end{equation}
as shown in Fig.~\ref{fig:kgm_XY} (d). In fact, the same tensor network has
been also obtained in a less straightforward way with the help of the
infinite summation, where the interactions between emergent variables can be
seen clearly\cite{Song_2023_2}. A direct comparison to the problematic
standard construction in Ref. \cite{Song_2023_2} demonstrates the importance
of encoding the emergent degree of freedom in the local tensors: besides the
proper Hamiltonian tessellation, the duality transformation is also
necessary to capture the essential physics of the chiralities.

In the framework of tensor networks, the entanglement entropy of the
fixed-point MPS for the 1D quantum correspondence exhibits singularity at
the critical temperatures, offering a sharp criterion to determine possible
phase transitions in the thermodynamic limit. As shown in Fig.~\ref%
{fig:kgmxy_ee}, by employing the tensor network method outlined in the
Appendix, the entanglement entropy $S_E$ develops only one sharp singularity
at the critical temperature $T_c\simeq0.075J$, indicating that a single BKT
phase transition takes place at a rather low temperature. The peak positions
are almost unchanged with different MPS bond dimensions ranging from $D=60$
to $120$. Thus, the transition temperature is determined with high
precision, which is in good agreement with theoretical predictions for the
unbinding temperature of $1/3$ vortex pairs\cite%
{Cherepanov_2001,Korshunov_2002,Song_2023_2}. The low-temperature phase of
the model can be interpreted as the presence of charge-6e superconductivity
(SC) in the absence of charge-2e SC\cite{Song_2023_2}.

\begin{figure}[tbp]
\centering
\includegraphics[width=\linewidth]{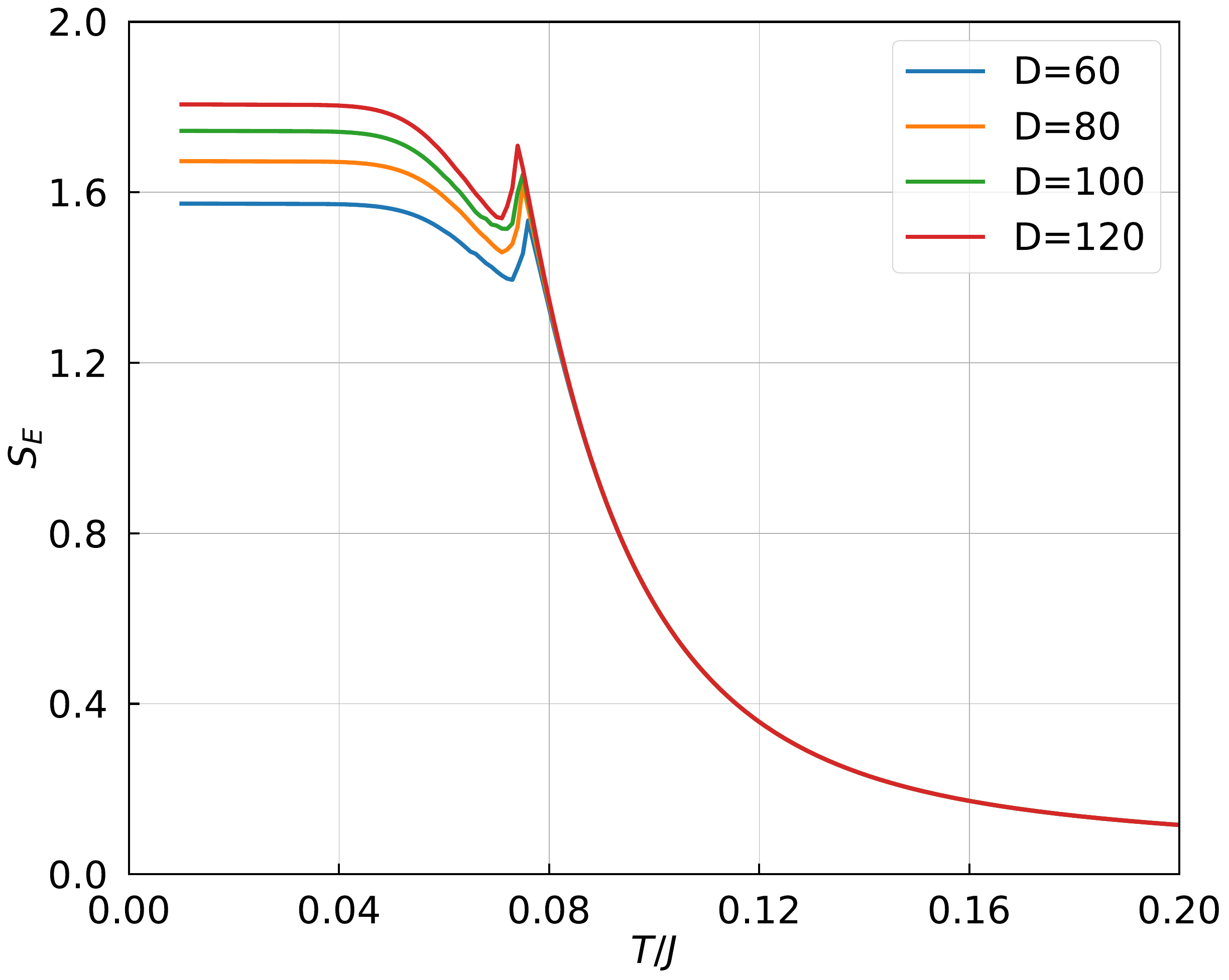}
\caption{The entanglement entropy as a function of temperature under
different MPS bond dimensions for the AF XY spin model on the kagome
lattice. }
\label{fig:kgmxy_ee}
\end{figure}

\subsection{Fully Frustrated XY spin model on a square lattice}

\begin{figure}[tbp]
\centering
\includegraphics[width=\linewidth]{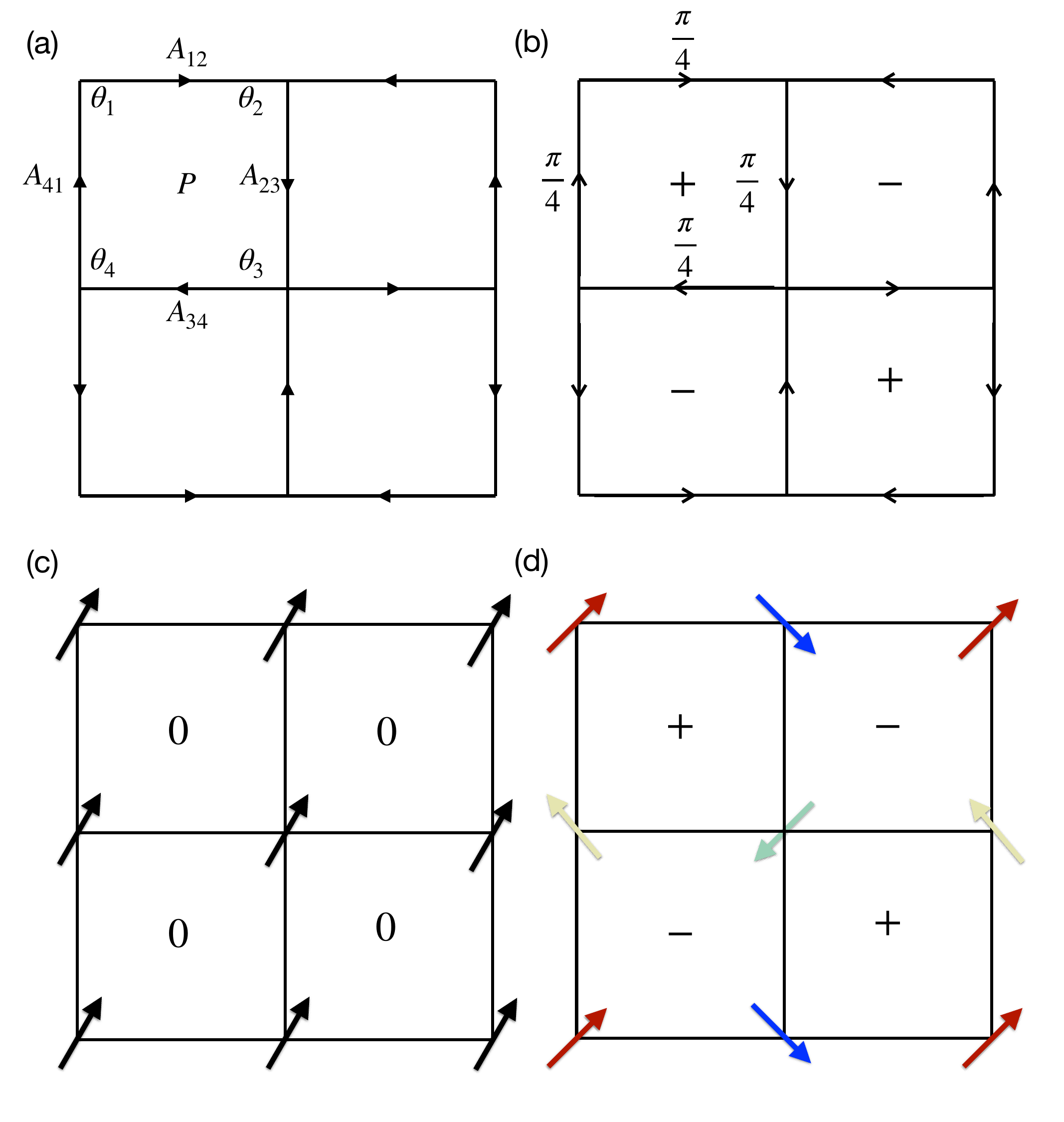}
\caption{ (a) The fully frustrated XY model on a square lattice. The arrows
on the links correspond to the gauge field ${A_{ij}}$ with the value of $\pm%
\frac{\protect\pi}{4}$. The sign of $A_{ij}$ is denoted by the direction of
the arrow. (b) The ground state of the FFXY model on a square lattice with a
checkboard pattern of chirality. (c) The ground state of the modified XY
model for $\frac{\protect\mu}{J}<\frac{1}{8}$. (d) The ground state of the
modified XY model for $\frac{\protect\mu}{J}>\frac{1}{8}$. The $0,\pm$ signs
correspond to the topological charges located on the centers of the
plaquettes.}
\label{fig:sqr_XY_cfg}
\end{figure}

The fully frustrated XY (FFXY) spin model on a 2D square lattice can be
defined with gauge fields on the lattice bonds
\begin{equation}
H=-J\sum_{\langle i,j\rangle }\cos (\theta _{i}-\theta _{j}-A_{ij}),
\end{equation}%
where the full frustration corresponds to the uniform gauge field of $%
A_{ij}=\pi /4$ on each bond of the square plaquettes. As displayed in Fig.~%
\ref{fig:sqr_XY_cfg} (b), the minimum of the Hamiltonian is obtained when
all gauge-invariant phase differences between nearest-neighbor spins $\phi
_{i,j}=\theta _{i}-\theta _{j}-A_{ij}$ equal to $\pm \pi /4$. The ground
state can be characterized by a checkerboard pattern of chiralities $\tau
=\pm 1$ defined by $\sum_{\square }\phi _{i,j}=\tau\pi $. Another degenerate
state can be obtained by switching the positive and negative chiralities.
Therefore, in addition to the $U(1)$ symmetry, the chiralities give rise to
an emergent $Z_{2}$ degeneracy of the ground state of the FFXY model on a
square lattice~\cite{Villain1977,Villain1977_2,Song_2022}.

\begin{figure*}[tbp]
\centering
\includegraphics[width=\linewidth]{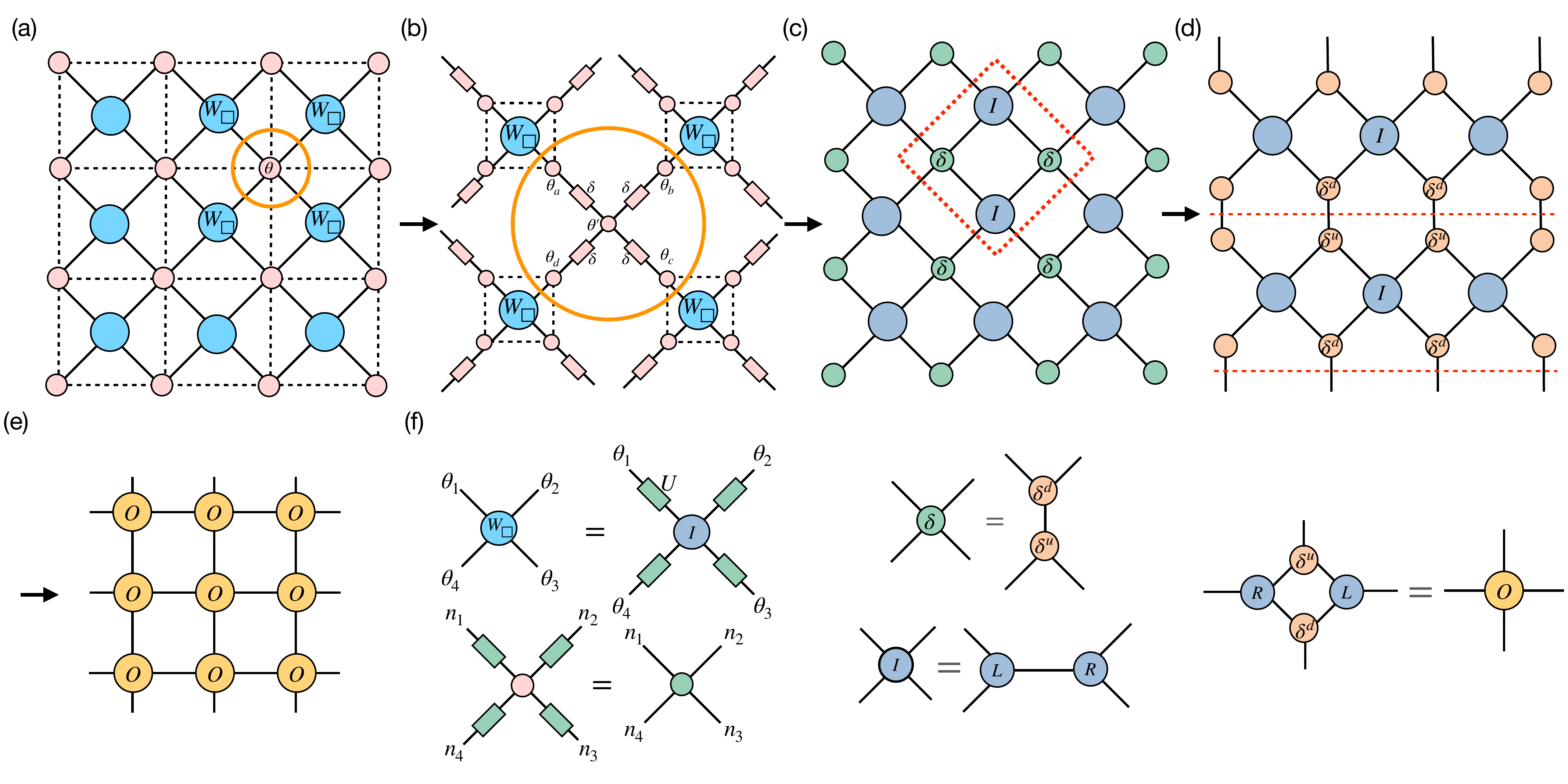}
\caption{ Tensor network representation of the FFXY model on a square
lattice. (a) The tensor network with continuous indices, where the $W$
tensors account for the Boltzmann weight on each square and the pink dot
tensor accounts for the integration of the shared $\protect\theta$ variables
among four plaquettes. The black dotted line denotes the original square
lattice. (b) The auxiliary spin $\protect\theta^{\prime }$ connecting the
copied spins of four nearby plaquettes, the $\protect\delta$ matrices
represent the Dirac delta functions. (c) The tensor network with discrete
indices obtained from Fourier transformations on the $W$ tensors and the
integrations on the $\protect\theta$ variables. (d) The row-to-row transfer
matrix built by splitting the $\protect\delta$ tensors vertically. (e) The
uniform tensor network representation composed of local tensor $O$. (f) The
details of the operations on the local tensors in the construction of the
tensor network. }
\label{fig:sqr_XY}
\end{figure*}

To obtain the tensor network representation of the partition function, we
first divide the global Hamiltonian into a tessellation of local Hamiltonian
on each square where the emergent variables live
\begin{equation}
H=\sum_{\square}H_{\square},
\end{equation}
and the local cluster of interactions is given by
\begin{equation}
H_{\square}=-\frac{J}{2}\sum_{\langle i,j\rangle\in{\square}
}\cos(\theta_i-\theta_j-A_{ij}).
\end{equation}

Then the tensor network can be expressed as a product of local Boltzmann
weights on each plaquette as shown in Fig.~\ref{fig:sqr_XY} (a)
\begin{equation}
Z=\prod_{i}\int\frac{d\theta_i}{2\pi} \prod_{\square} W_{\square}
\label{eqn:ffsXY_rep}
\end{equation}
where $W_p(\theta_1,\theta_2,\theta_3,\theta_4)=\mathrm{e}^{-\beta
H_{\square}}$ is a four-legged tensor with a continuous $U(1)$ indices.

Different from the corner-shared case of the kagome lattice, particular
attention should be paid to the split of the shared spins among four square
plaquettes. To avoid the formation of loops of the Dirac delta functions
among four $W$ tensors
\begin{equation}
\delta (\theta _{a}-\theta _{b})\delta (\theta _{b}-\theta _{c})\delta
(\theta _{c}-\theta _{d})\delta (\theta _{d}-\theta _{a})  \notag
\end{equation}%
with $\theta _{a}$, $\theta _{b}$, $\theta _{c}$ and $\theta _{d}$
representing the four replicas of the shared spin, we put an auxiliary spin $%
\theta _{i}^{\prime }$ connecting to the shared spins
\begin{equation}
\delta (\theta _{a}-\theta ^{\prime })\delta (\theta _{b}-\theta ^{\prime
})\delta (\theta _{c}-\theta ^{\prime })\delta (\theta _{d}-\theta ^{\prime
})  \notag  \label{eq:Dirac_star}
\end{equation}%
in a star shape as shown in Fig.~\ref{fig:sqr_XY} (b). Then we transform the
local tensor $W_{p}$ to the discrete basis%
\begin{eqnarray}
I_{n_{1},n_{2},n_{3},n_{4}} &=&\prod_{i=1}^{4}\int \frac{d\theta _{i}}{2\pi }%
W_{\Delta }(\theta _{1},\theta _{2},\theta _{3},\theta _{4})  \notag \\
&&\cdot U_{n_{1}}(\theta _{1})U_{n_{2}}(\theta _{2})U_{n_{3}}(\theta
_{3})U_{n_{4}}(\theta _{3}),
\end{eqnarray}
where $U_{n}(\theta )$ are the Fourier basis defined in \eqref{eq:Fourier}.

\begin{figure}[tbp]
\centering
\includegraphics[width=\linewidth]{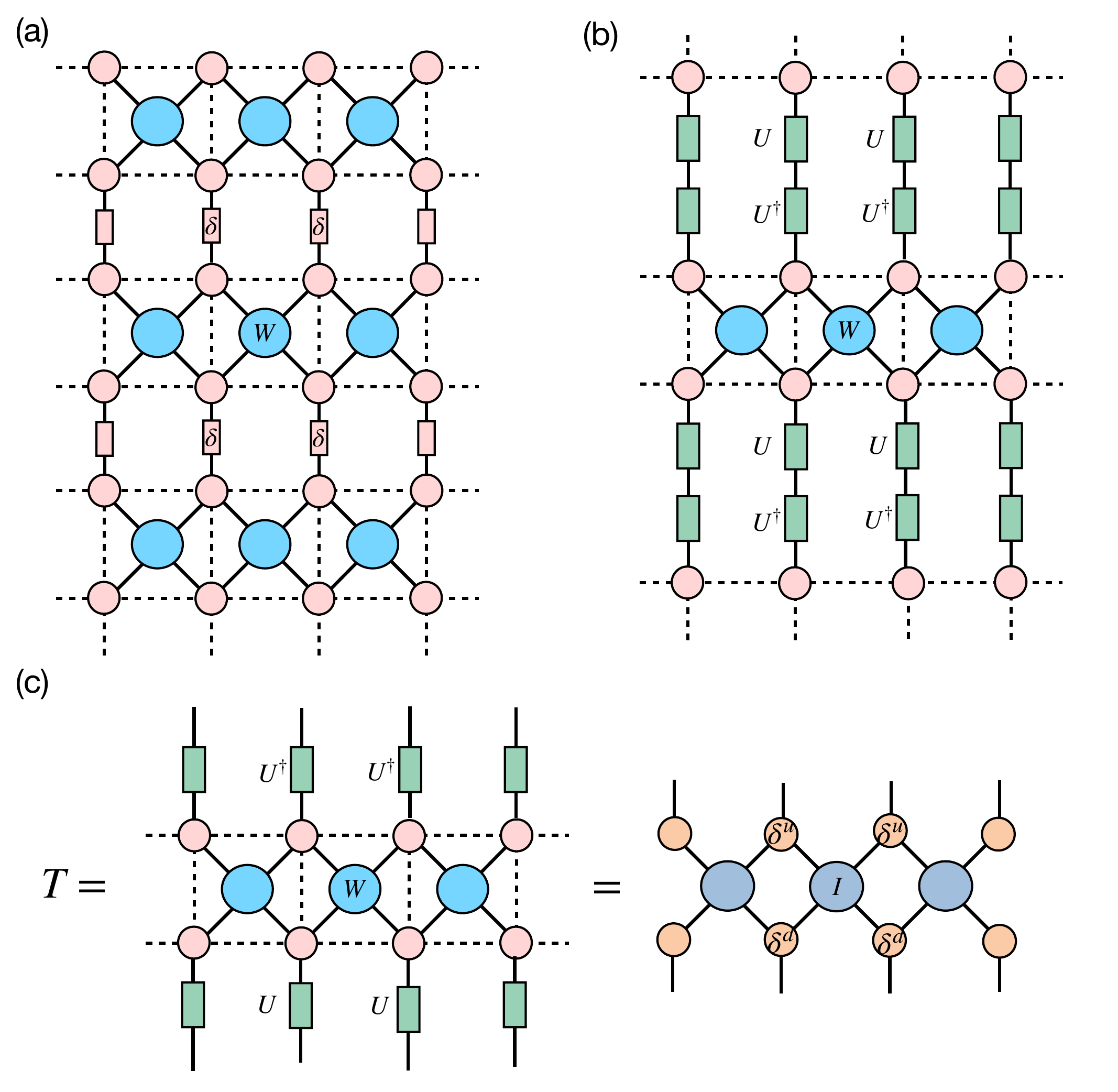}
\caption{ (a) Tensor network representation of the partition function with
the split of $U (1)$ phase variables vertically. The $\protect\delta$
matrices represent the Dirac delta functions. (b) The decomposition of the
Dirac delta function into $U$ matrix and $U^{\dagger}$ matrices. (c) The
row-to-row transfer matrix on the continuous basis and the discrete basis,
respectively. }
\label{fig:sqr_XY_vsp}
\end{figure}

As shown in Fig.~\ref{fig:sqr_XY} (f), the constraint of the star-shaped
Dirac delta functions \eqref{eq:Dirac_star} can be reduced to a four-legged
Kronecker delta tensor via
\begin{equation}
\delta_{n_1+n_2+n_3+n_4,0}=\int\frac{d\theta^{\prime }}{2\pi}%
U_{n_1}(\theta^{\prime })U_{n_2}(\theta^{\prime })U_{n_3}(\theta^{\prime
})U_{n_4}(\theta^{\prime })  \notag  \label{eq:conserv_delta}
\end{equation}
characterizing the conservation law of $U(1)$ charges. As a result, we get
the tensor network representation composed of local tensors of discrete
indices as displayed in Fig.~\ref{fig:sqr_XY} (c).

Furthermore, the $\delta $ tensors are split vertically as shown in Fig.~\ref%
{fig:sqr_XY} (d),
\begin{equation}
\delta _{n_{1}+n_{2}+n_{3}+n_{4},0}=\sum_{n_{5}}\delta
_{n_{1}+n_{2}-n_{5},0}^{u}\delta _{n_{3}+n_{4}+n_{5},0}^{d}
\label{eq:vsp_delta}
\end{equation}
and the $I$ tensors are decomposed horizontally by SVD
\begin{equation}
I_{n_{1},n_{2},n_{3},n_{4}}=%
\sum_{n_{5}}L_{n_{1},n_{2},n_{5}}R_{n_{5},n_{3},n_{4}}
\end{equation}
as displayed in Fig.~\ref{fig:sqr_XY} (f). Finally, the regular local tensor
$O$ in the uniform tensor network of Fig.~\ref{fig:sqr_XY} (e) is obtained
by grouping the relevant component tensors.

One might rotate the network in Fig.~\ref{fig:sqr_XY} (c) by $45$ degrees
and group the local tensors in the red dotted line to directly make up a
four-legged translation-invariant local tensor. However, the standard
contraction algorithms fail to converge under this construction because the
linear transfer matrix is non-Hermitian. Another key insight is that such a
construction does not take into account the checkerboard-like ground state
configurations, where only two chiralities are included in the transitional
unit.

Actually, although the procedure of the construction is different, the
tensor network in Fig.~\ref{fig:sqr_XY} turns out to share the same transfer
matrix as the one obtained in the Ref. \cite{Song_2022}. To prove it, we
split the $U(1)$ spins in the vertical direction using the relation
\begin{eqnarray}
\int d\theta_if(\theta_i)=\iint d\theta_id\theta_i^{\prime
}\delta(\theta_i-\theta_i^{\prime })f(\theta_i^{\prime }),
\end{eqnarray}
where the spin $\theta_i^{\prime }$ is a copy of spin $\theta_i$ connected
by the Dirac delta function as shown in Fig.~\ref{fig:sqr_XY_vsp} (a). The
delta tensor on a link can be further decomposed by the Fourier basis
\begin{equation}
\delta(\theta-\theta^{\prime })=\frac{1}{2\pi}\sum_n
U_n^{\dagger}(\theta^{\prime })U_n(\theta),
\end{equation}
as displayed in Fig.~\ref{fig:sqr_XY_vsp} (b). Now we can define the row to
row transfer matrix as three stripes of $U$, $W$ and $U^\dagger$ tensors as
shown in Fig.~\ref{fig:sqr_XY_vsp} (c). It is easy to see that the transfer
matrix is Hermitian just like the one constructed in Ref.\cite{Song_2022}
since the $W$ tensors are real and symmetric. Using the Fourier
transformation again, we get the same $I$ and $\delta$ tensors in Fig.~\ref%
{fig:sqr_XY_vsp} (c) as those displayed in Fig.~\ref{fig:sqr_XY} (d).

\begin{figure}[tbp]
\centering
\includegraphics[width=\linewidth]{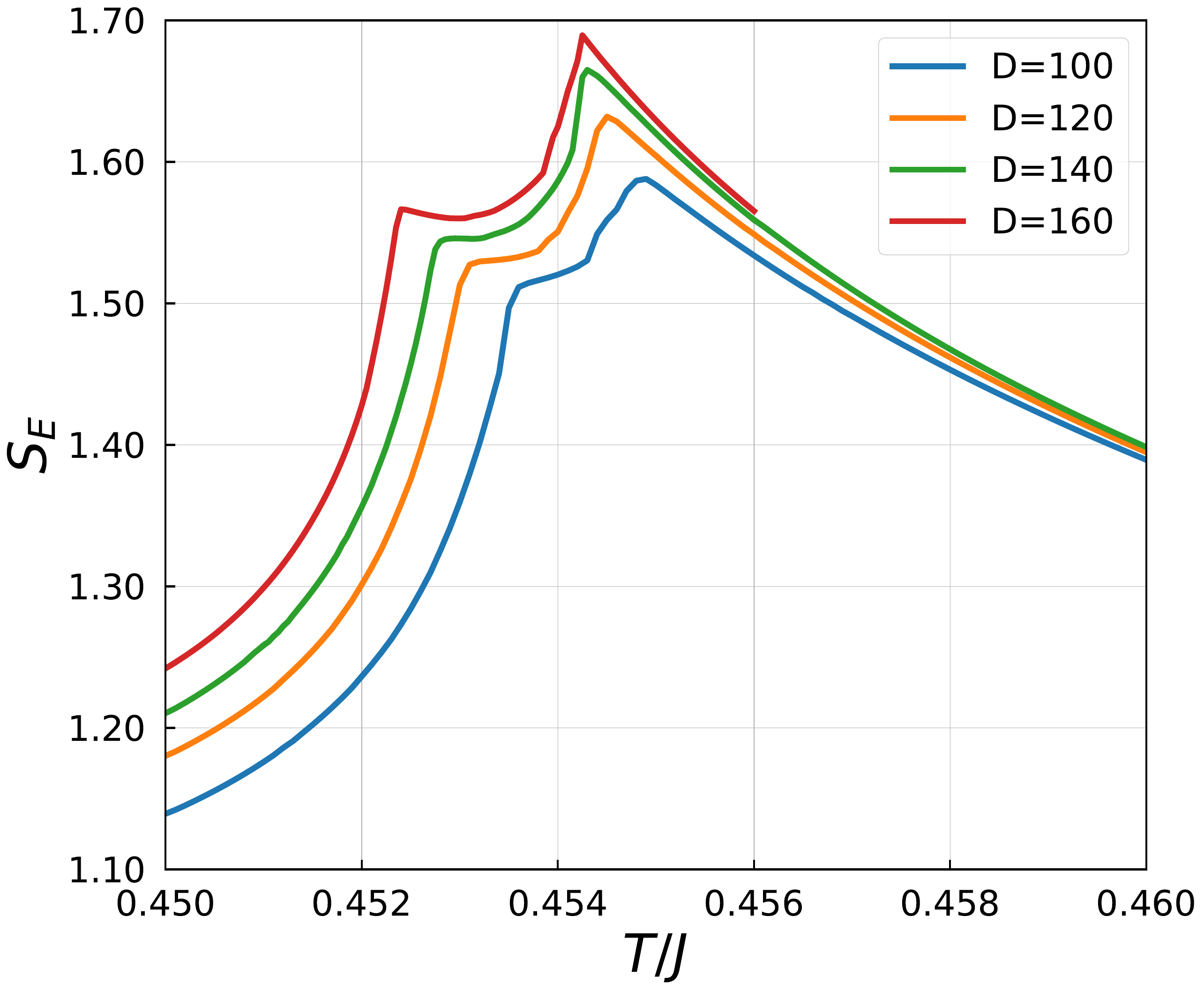}
\caption{ The entanglement entropy for the FFXY model on the square lattice
develops two singularities indicating the existence of two phase transitions
with the increasing of MPS bond dimensions.}
\label{fig:ffxy_sq_ee}
\end{figure}

Once the proper tensor network representation is obtained, the numerical
calculations can be efficiently performed as illustrated in the Appendix. As
shown in Fig.~\ref{fig:ffxy_sq_ee}, the entanglement entropy $S_E$ develops
two sharp singularities at two critical temperatures $T_{c1}$ and $T_{c2}$,
which strongly indicates the existence of two phase transitions at two
different temperatures. As the singularity positions vary with the MPS bond
dimension $D$, the critical temperatures $T_{c1}$ and $T_{c2}$ can be
determined precisely by extrapolating the bond dimension $D$ to infinite.
Moreover, we find that the critical temperatures $T_{c1}$ and $T_{c2}$
exhibit different scaling behaviors in the linear extrapolation, implying
that the two phase transitions belong to different kinds of universality
classes. The lower transition temperature $T_{c1}$ varies linearly on the
inverse square of the logarithm of the bond dimension, while the higher
transition temperature $T_{c2}$ has a linear variance with the inverse bond
dimension. The different scaling behavior agrees well with the different
critical behavior of the BKT and 2D Ising universality classes\cite%
{Song_2022}.

\begin{figure*}[t]
\centering
\includegraphics[width=\linewidth]{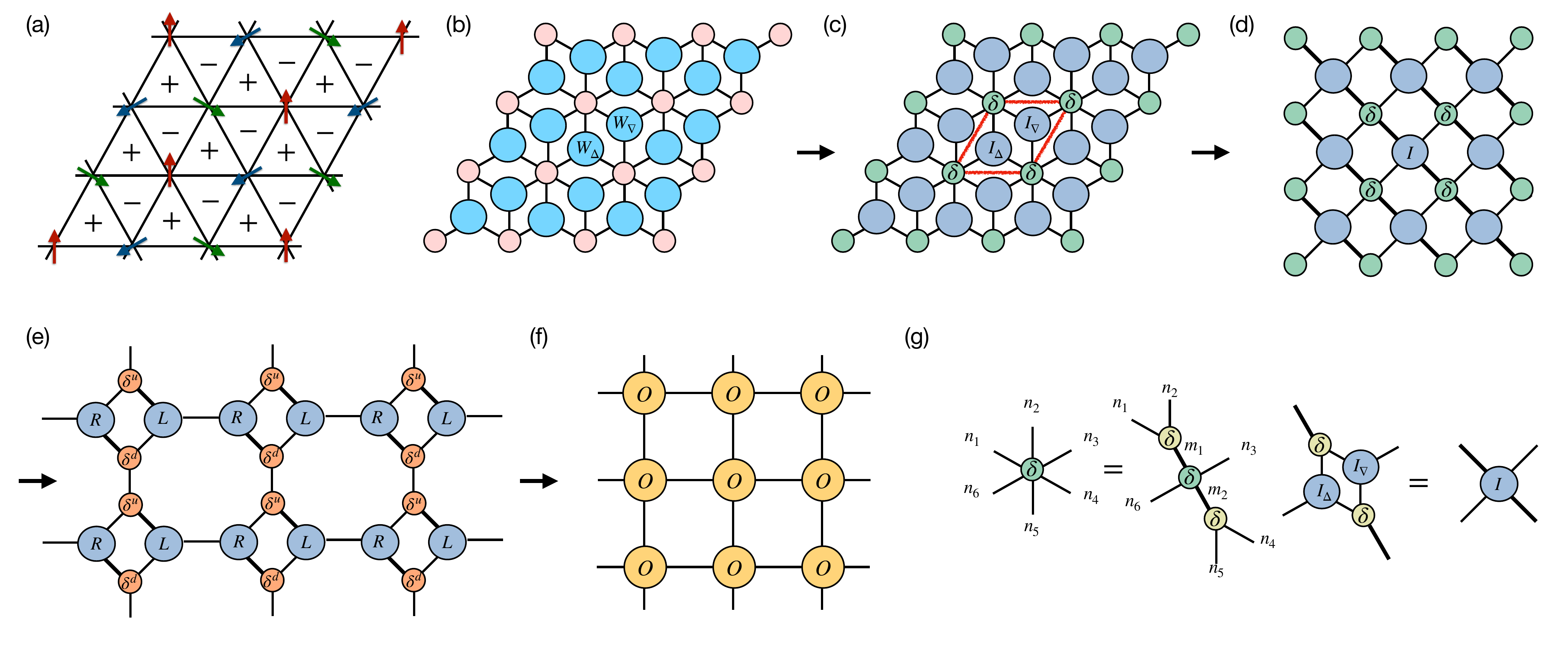}
\caption{ The tensor network representation of the FFXY model on a
triangular lattice. (a) One of the ground state spin configurations. The
chiralities denoted by plus and minus signs on the centers of the triangular
plaquettes form an AF pattern. (b) The tensor network with continuous
indices. (c) The tensor network is transformed onto a discrete basis through
the Fourier transformation. A parallelogram unit cell is circled in the red
line. (d) The $I$ tensors is constructed by grouping a pair of $I_{\protect%
\nabla}$ and $I_{\Delta}$ tensors. (e) The vertical split of the $\protect%
\delta$ tensors into $\protect\delta^d$ and $\protect\delta^u$ and the
horizontal split of the $I$ tensors into $L$ and $R$. (f) The tensor network
representation composed of uniform local $O$ tensors. (g) Details of the
transformations of local tensors.}
\label{fig:tri_XY}
\end{figure*}

\subsection{AF XY spin model on a triangular lattice}

The frustrated XY spin model on a triangular lattice under a fixed gauge
condition of $A_{ij}=\pi $ on each triangular plaquette can be transformed
into an AF XY spin model. As shown in Fig.~\ref{fig:tri_XY} (a), the angle
between each pair of the nearest-neighbor spins should be $\pm 2\pi /3$ to
achieve the minimum of the ground state energy. Like the FFXY model on the
square lattice, the elementary triangular plaquettes can be characterized by
alternating chiralities of $\tau =\pm 1$. The translation-invariant unit of
the spin configuration forms a $3\times 3$ cluster larger than the original
lattice.

The tensor network can be constructed in the same way as the FFXY spin model
on the square lattice. First, we decompose the Hamiltonian into local terms
on each triangle
\begin{equation}
H=\sum_{p}H_p, \hspace{0.5cm} H_p=\frac{J}{2}\sum_{\langle i,j\rangle\in
p}\cos(\theta_i-\theta_j).
\end{equation}
The partition function can be expressed as a product of local Boltzmann
weights
\begin{equation}
Z=\prod_{i}\int\frac{d\theta_i}{2\pi}\prod_{p}W_p,
\end{equation}
where $W_p=\mathrm{e}^{-\beta H_p}$ defined on the centers of the triangles
are three-legged tensors sharing the same $U(1)$ spin at the joint corners
as shown in Fig.~\ref{fig:tri_XY} (b).

Then the $W$ tensors and the Dirac delta functions are transformed onto a
discrete basis by the Fourier transformations, as displayed in Fig.~\ref%
{fig:tri_XY} (c). To achieve a transition-invariant unit, we take a
parallelogram cell circled by the red line and reorganize the local tensors
within it. As shown in Fig.~\ref{fig:tri_XY} (g), the six-legged delta
tensor is decomposed into three smaller delta tensors%
\begin{eqnarray}
&&\delta _{n_{1}+n_{2}+n_{3}+n_{4}+n_{5}+n_{6},0}  \notag \\
&=&\sum_{m_{1},m_{2}}\delta _{n_{1}+n_{2},m_{1}}\delta
_{m_{1}+n_{3}+m_{2}+n_{6},0}\delta _{n_{4}+n_{5},m_{2}},  \notag
\end{eqnarray}%
where the bond dimension of the $m$-indexed leg is bigger than the $n$%
-indexed leg denoted by a thicker line. At the same time, a pair of $%
I_{\Delta }$ and $I_{\nabla }$ tensors are grouped together into a
four-legged $I$ tensor%
\begin{eqnarray*}
I_{n_{1},m_{2},n_{3},m_{4}}&=&\sum_{n_{2},n_{4},n_{5},n_{6}}\delta
_{n_{2}+n_{4},m_{2}} \notag \\
&&(I_{\Delta })_{n_{1},n_{2},n_{5}}\delta_{n_{5}+n_{6},m_{4}}(I_{\nabla })_{n_{3},n_{4},n_{6}},
\end{eqnarray*}
and the tensor network is transformed to a relatively structured form in
Fig.~\ref{fig:tri_XY} (d). Following the same procedure of a vertical split
of the $\delta $ tensors and a horizontal split of the $I$ tensors, we
obtain the uniform tensor network in Fig.~\ref{fig:tri_XY} (f).

Note that the Fourier transformation must be performed on each triangular
plaquette first to ensure the emergence of the dual variables. Otherwise, if
we directly choose a parallelogram including a pair of neighboring triangles
and then build the tensor network based on the local Boltzmann weight of
\begin{eqnarray}
&& W_{\parallelogramm}(\theta_1,\theta_2,\theta_3,\theta_4)  \notag \\
&&=\exp\Big\{-\frac{\beta J}{2}[\cos(\theta_1-\theta_2)+\cos(\theta_2-%
\theta_3)  \notag \\
&&+\cos(\theta_3-\theta_4)+\cos(\theta_4-\theta_1)+2\cos(\theta_1-\theta_3)]%
\Big\},  \notag
\end{eqnarray}
the infinite contraction of the tensor network will not give the right
results. The reason is that the construction of local tensors with a finite
bond cut-off can be regarded as a coarse-grained procedure that should be
performed exactly on the clusters of spin corresponding to the emergent
degrees of freedom.

\begin{figure}[tbp]
\centering
\includegraphics[width=\linewidth]{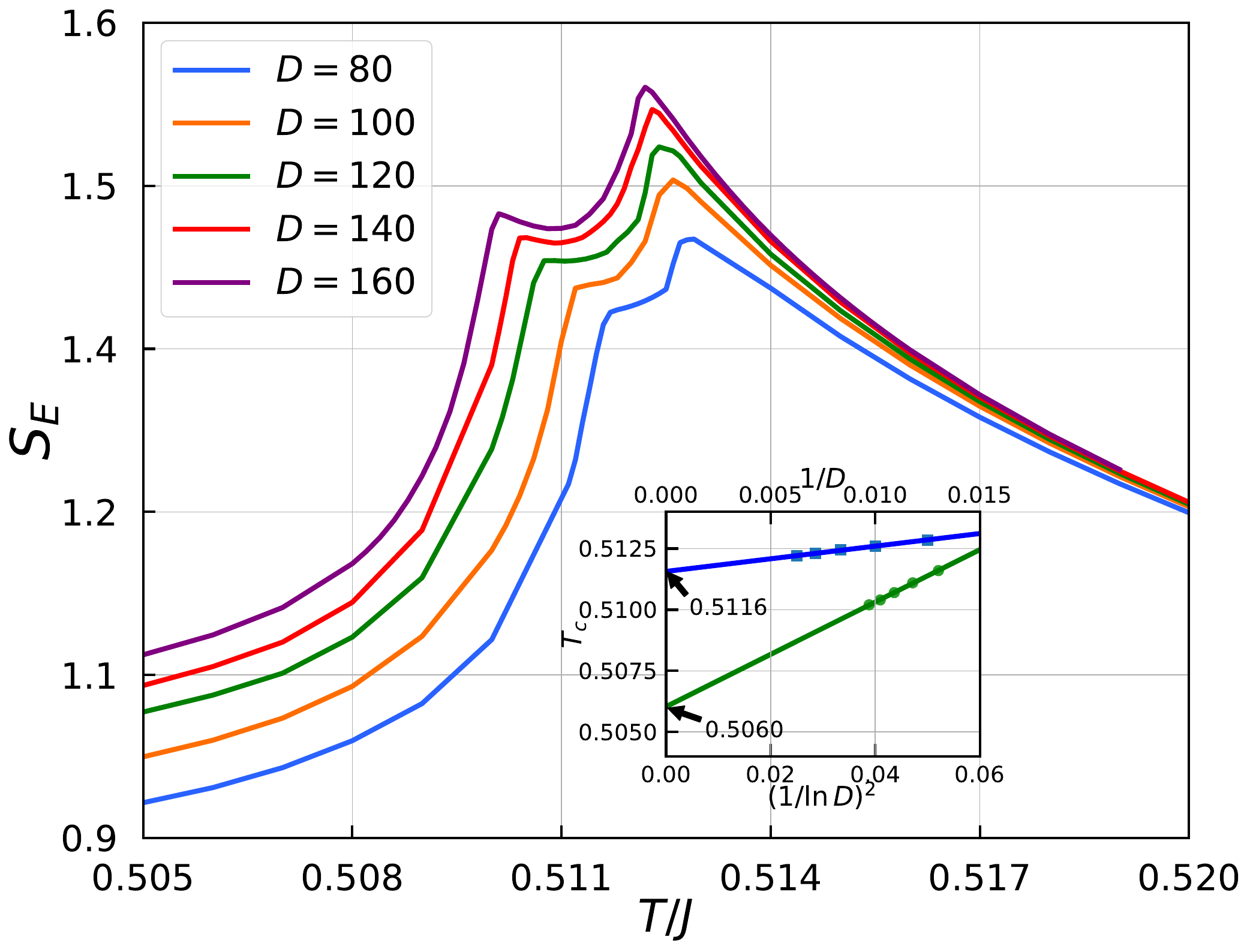}
\caption{ For the AF XY model on the triangular lattice, the entanglement
entropy as a function of temperature develops two peaks when the MPS bond
dimension $D$ is increased. Inset: The singularity temperatures $T_{c1}$ and
$T_{c2}$ of the entanglement entropy fitted for MPS bond dimensions from $%
D=80$ to $160$.}
\label{fig:afxytr_ee}
\end{figure}

As shown in Fig.~\ref{fig:afxytr_ee}, the entanglement entropy $S_{E}$ also
develops two sharp singularities at two critical temperatures $T_{c1}$ and $%
T_{c2}$, and the critical temperatures have the same scaling behavior as the
FFXY model on the square lattice. From the linear extrapolation, the
critical temperatures are estimated to be $T_{c1}\simeq 0.5060J$ and $%
T_{c2}\simeq 0.5116J$. The critical temperature $T_{c1}$ agrees well with
previous Mont Carlo results \cite{Obuchi_2012} obtained by BKT fitting and $%
T_{c2}$ is slightly lower than a recent estimation \cite{Obuchi_2012,Lv_2013}
of $T_{c2}\simeq 0.512J$.

\begin{figure}[tbp]
\centering
\includegraphics[width=\linewidth]{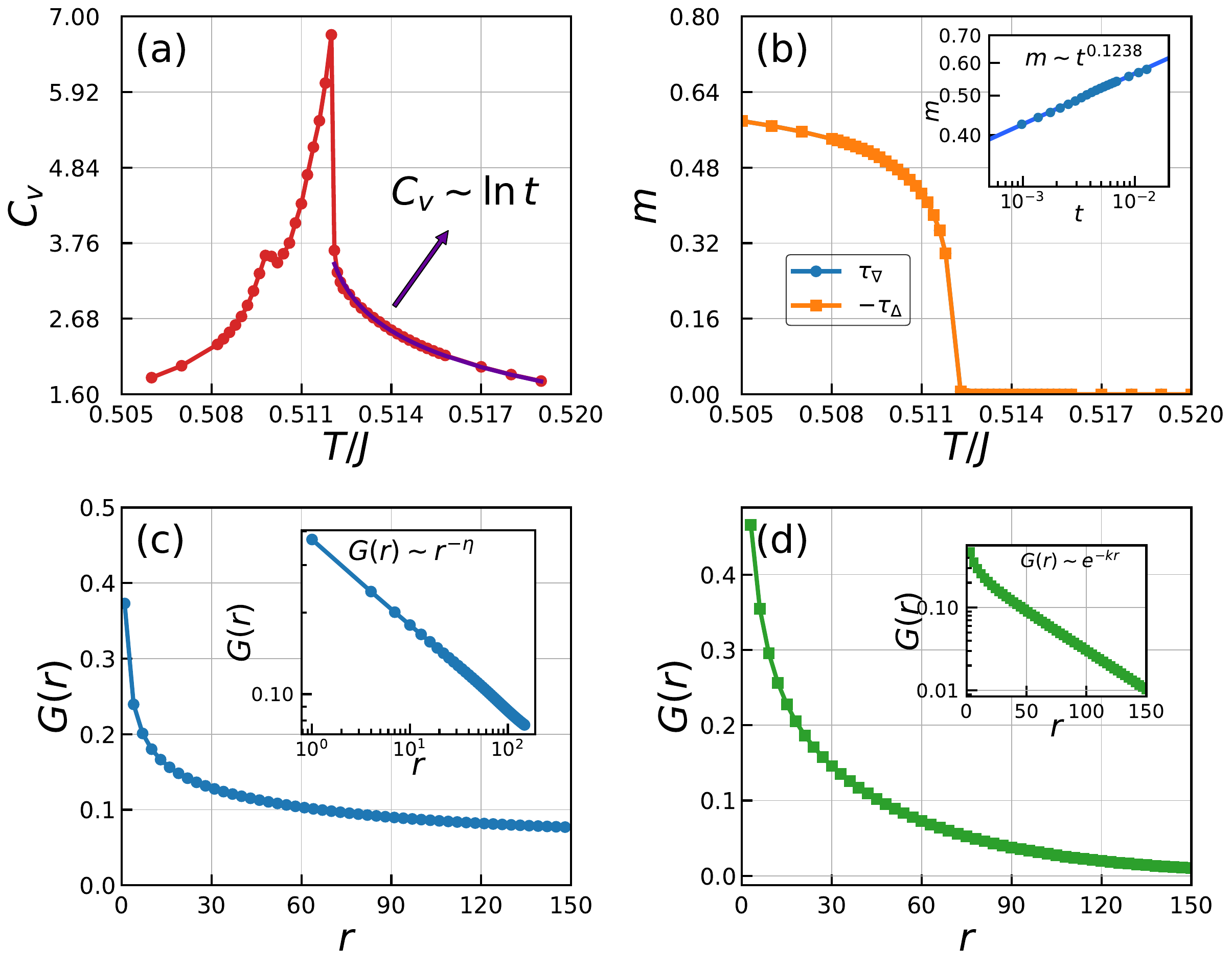}
\caption{ (a) The specific heat shows a small bump around $T_{c1}$ but a
logarithmic divergence at $T_{c2}$. (b) The symmetry breaking of chirality
at $T_{c2}$. The inset is the fitting of the Ising critical exponent. (c)
The spin-spin correlation function shows a power-law decay below $T_{c1}$
(d) The spin-spin correlation function shows an exponential decay above $%
T_{c1}$.}
\label{fig:afxytr}
\end{figure}

The properties of the two distinct phase transitions can be further
elucidated through the thermodynamic quantities. The results of the specific
heat are presented in Fig.~\ref{fig:afxytr} (a). Around the critical
temperature $T_{c1}$, the specific heat exhibits a small bump, indicating a
higher-order continuous phase transition. By comparison, the specific heat
displays a sharp divergence at $T_{c2}$, implying a second-order phase
transition. For the high-temperature side $T>T_{c2}$, the specific heat can
be fitted well by the logarithmic behavior of a second-order Ising
transition. The specific heat between $T_{c1}$ and $T_{c2}$ does not fit
well with the logarithmic form due to the close proximity of the two
transitions. The breaking of $Z_{2}$ symmetry at $T_{c2}$ can be
demonstrated by the expectation values of the chiralities. As shown in Fig.~%
\ref{fig:afxytr} (b), below the critical temperature $T_{c2}$, the chiral
order parameter
\begin{equation}
m=\frac{1}{N}\sum_p (-1)^{x+y}\tau_p  
\label{eq:chiral_op}
\end{equation}
associated with the chiral degrees of freedom establishes a non-zero value,
corresponding to the checkerboard pattern of chirality on upward and
downward triangles. When approaching the critical temperature $T_{c2}$ from
the low-temperature side, the order parameter vanishes continuously as $m
\sim t^{\beta }$ with $t=1-T/T_{c2}$. The critical exponent $\beta \simeq
0.1238$ is in good agreement with the critical exponent $\beta =1/8$ for the
2D Ising universality class.

The nature of the phase transition at $T_{c1}$ can be revealed in the change
of the behavior of the spin-spin correlation functions defined as
\begin{equation}
G(r)= \langle \cos(\theta_{i}-\theta_{i+r}) \rangle.  \label{eq:G}
\end{equation}
A comparison of correlation functions below and above $T_{c1}$ is displayed
in Fig.~\ref{fig:afxytr}(c) and (d). Below $T_{c1}$, the spin-spin
correlation function exhibits a power-law decay, implying a close binding
between vortices and anti-vortices. In contrast, for $T>T_{c1}$ the
correlation function displays an exponential decay, indicating the
destruction of phase coherence between vortices due to the unbinding of
vortex pairs. Thus, the phase transition at $T_{c1}$ belongs to the
universality class of the BKT transition.

\subsection{Modified XY model on a square lattice}

The unified tensor network methods can be employed in the study of
frustrated spin models with more complex interactions. One such model is the
modified XY model defined on a 2D square lattice \cite{Maccari_2020,Maccari_2023}
\begin{equation}
H=-J\sum_{\langle i,j\rangle }\cos (\theta _{i}-\theta _{j})-\mu
\sum_{p}\tau _{p}^{2},
\end{equation}
where the first term is the original XY model of ferromagnetic coupling $J>0$%
, and the second term tunes the vortex fugacity through the chemical
potential $\mu $. The spin current circulating around each single square
plaquette is defined as
\begin{equation}
\tau _{p}=\sum_{\langle i,j\rangle \in p}\sin (\theta _{i}-\theta _{j}).
\notag
\end{equation}%

It is well-known that the original XY spin model can be mapped into an
interacting Coulomb gas with a vortex-core energy fixed in the low-density
limit \cite{J.M.Kosterlitz_1973, J.M.Kosterlitz_1974}. And the underlying
physics at large vortex density is of general interest both theoretically
and experimentally. In the area of theoretical investigations, the possible
extension of BKT theory under a large vortex fugacity was discussed, where
non-BKT behavior and the occurrence of first-order transition were proposed  \cite{Minnhagen_1985,Minnhagen_1985_2,Minnhagen_1987,Zhang_GM_1993}. Actually
a generalization of 2D XY spin model with a "crossed-product" operator acting
on the plaquettes had been introduced to adjust the core energy of the
vortices \cite{Swendsen_1982}. Subsequently, the numerical explorations of
a Coulomb gas model on the square and triangular lattices as well as in the
continuous limit showed a rich phase diagram with novel critical behaviors of
an ordered-charge lattice \cite{lee_1990, lee_1991,lidmar_1997}. Moreover,
the similar physics has been investigated in 3D XY spin models, where a term
acting on the plaquette was introduced to regulate the energy of vortex strings
\cite{kohring_1986,shenoy_1990}.

The experiments in superconducting thin films revealed a significant
deviation of the vortex-core energy from the predictions in the original XY
model \cite{kamlapureMeasurementMagneticPenetration2010}. It was found that
an accurate consideration of the vortex-core energy is of great importance
for the experimental identification of the BKT transition \cite{Mondal_2011}%
. Apart from the widely known superfluid phase and normal phase, the
measurement of the third sound mode in $^{4}$He thin films suggested the
existence of a new phase \cite{Chen_1992}. To provide a theoretical
explanation for this phenomenon, researchers have proposed a fascinating
concept involving the formation of a lattice composed of vortices and
anti-vortices, with a remarkably low vortex core energy \cite%
{Zhang_SC_1993,Gabay_1993}. The existence of vortex-antivortex lattice has
also been proposed in other systems such as ultra-cold atoms \cite%
{Botelho_2006} and polariton fluids \cite{Hivert_2014}.

To understand the role of the modified interaction term $\tau _{p}$, we can
make a simple analysis of the ground state. The ground state structure can
be determined by the ratio of $\mu /J$ tuning the spin currents in the
system which effectively modulates the vortex fugacity. As illustrated in
Fig.~\ref{fig:sqr_XY_cfg}(c)-(d), when $\mu /J<1/8$, the ground state is
identical to that at $\mu =0$, corresponding to the ground state of the
original XY model where all spins align parallel to each other. As we
further increase the chemical potential to $\mu /J>1/8$, the ground state is
characterized by maximizing $\tau _{p}$ on each plaquette, resulting in a
phase difference of $\phi _{12}=\phi _{23}=\phi _{34}=\phi _{41}=\pm \pi /2$%
. This ground state has the same ground state degeneracy as the FFXY spin
model on a square lattice. From the perspective of vorticity, the ground
state at $\mu /J<1/8$ has zero vorticity at each plaquette termed as the
vortex vacuum state, whereas the ground state at $\mu /J>1/8$ has a
checkerboard pattern of vorticity equal to $\pm 1$ called the
vortex-antivortex crystal. Hence, the zero-temperature ground state
structure of the modified XY spin model is analogous to the 2D dense coulomb
gas on the square lattice \cite{lee_1990}.

\begin{figure}[t]
\centering
\includegraphics[width=\linewidth]{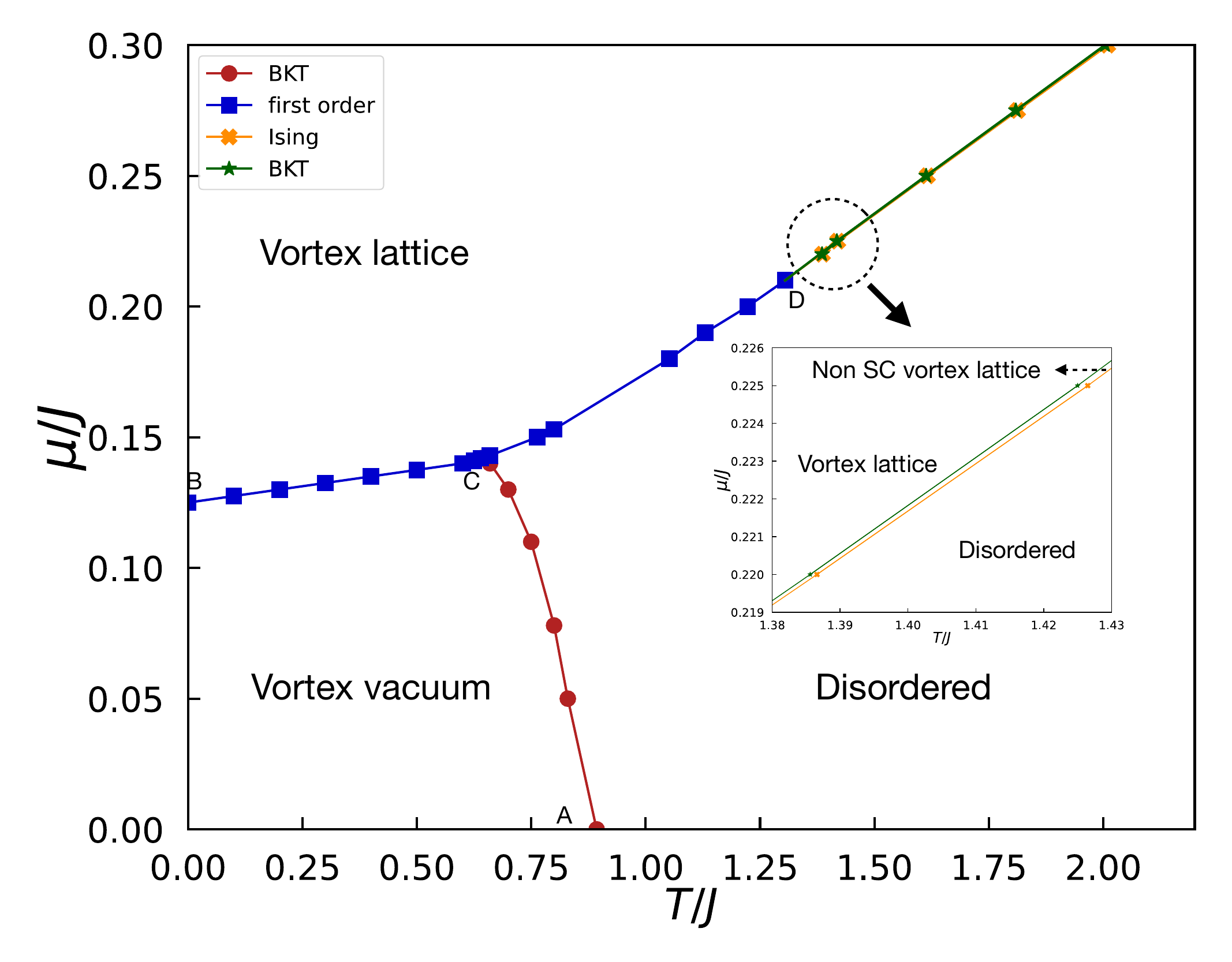}
\caption{ The global phase diagram of the modified XY spin model. The BKT
transition point A of the original XY spin model is determined as $(0.893,0)$%
. The exact solvable point B between the vortex vacuum phase and vortex
lattice phase at zero temperature is given by $(0.0,0.125)$. As the
temperature increases, depending on the chemical potential, the vortex
lattice can melt through three possible way. Below the point C at $%
(0.64,0.142)$, the vortex lattice experiences a first-order transition into
the vortex vacuum phase and then undergoes a BKT transition into the
disordered phase, while the CD line is the first-order transition. The point
D is a tricritical point determined as $\protect\mu/J\simeq 0.21$. Above
this point D, the first order transition line is separated into two
extremely close transition lines, belonging to the BKT transition and Ising
transition, respectively. Inset shows the enlarged results around the point $%
D$. }
\label{fig:mdfXY}
\end{figure}

The square term $\tau_p^2$ gives rise to multiple types of interaction
including the nearest-neighbor interactions, next-nearest-neighbor interactions,
and four-body interactions. Although it seems difficult to treat the
four-body interactions, there is still a well-defined vorticity on each
plaquette from the viewpoint of emergent degrees of freedom. Therefore we
can choose each square plaquette as an elementary cluster and replace the $%
H_p$ and $W_p$ by
\begin{equation}
H_p=-\frac{J}{2}\sum_{\langle i,j\rangle \in p }\cos(\theta_i-\theta_j)-\mu
\tau_p^2,\quad W_p=\mathrm{e}^{-\beta H_p}.
\end{equation}
Then the tensor network of the partition function can be constructed
following the procedure outlined in Fig.~\ref{fig:sqr_XY}. The singular
behavior of the entanglement entropy corresponding to the 1D transfer
operator offers a sharp criterion to determine all possible phase
transitions in the thermodynamic limit and the complete phase diagram is
thus determined as presented in Fig.~\ref{fig:mdfXY}.

\begin{figure}[tbp]
\centering
\includegraphics[width=\linewidth]{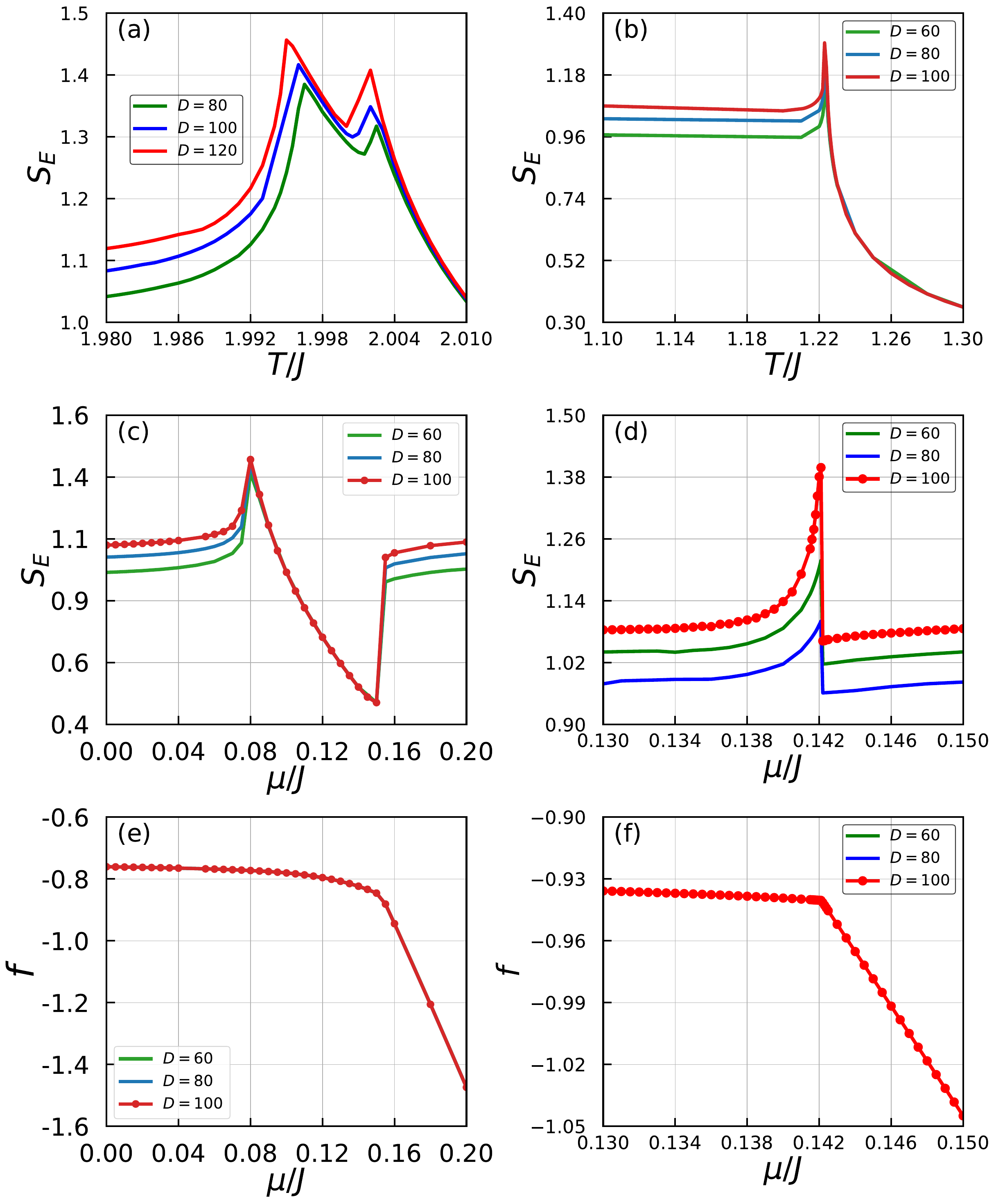}
\caption{ (a) The entanglement entropy as a function of temperature along $%
\protect\mu=0.3J$. (b) The entanglement entropy as a function of temperature
along $\protect\mu\simeq0.20J$. (c) The entanglement entropy as a function
of chemical potential along $T=0.8J$. (d) The entanglement entropy as a
function of chemical potential along $T=0.64J$. (e) The free energy density
as a function of chemical potential along $T=0.8J$. (f) The free energy
density as a function of chemical potential along $T=0.64J$.}
\label{fig:mdfxy_EE}
\end{figure}

\begin{figure}[tbp]
\centering
\includegraphics[width=\linewidth]{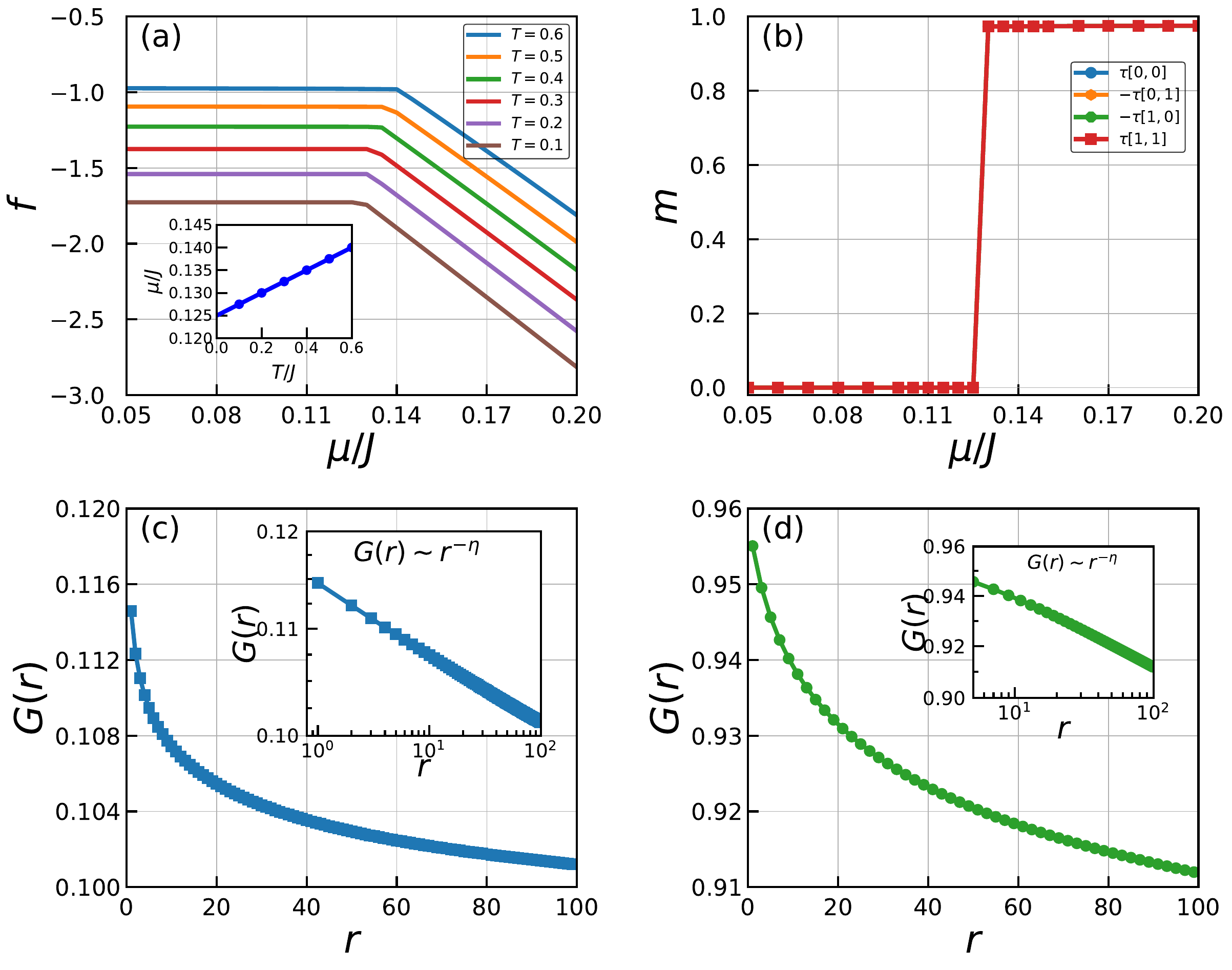}
\caption{ (a) The free energy density as a function of chemical potential at
different temperatures. Inset: linear extrapolation of critical chemical
potential as a function of temperature. (b) The checkboard-like chirality
pattern along $T=0.1J$. (c) The spin-spin correlation function shows an
exponential decay at $T\simeq0.1J$ and $\protect\mu\simeq0.1J$ in the
vortex-vacuum phase. (d) The spin-spin correlation function shows a
power-law decay at $T\simeq0.1J$ and $\protect\mu\simeq0.15J$ in the
vortex-lattice phase. }
\label{fig:mdfxy_low_T}
\end{figure}

In the upper plane of the phase diagram, the entanglement entropy along the
chemical potential $\mu=0.3J$ is displayed in Fig.~\ref{fig:mdfxy_EE} (a).
There exist two distinct peaks, corresponding to the BKT and Ising
transition, respectively. These two phase transitions are extremely close to
each other as shown by the zoomed inset in Fig.~\ref{fig:mdfXY}. Upon
further reducing the chemical potential to $\mu\simeq0.20J$, two separated
peaks merge into a single peak, as displayed in Fig.~\ref{fig:mdfxy_EE} (b).
The merging point is denoted as the point $D$ in the global phase diagram.
The low-temperature phase with large $\mu$ is called the vortex-lattice phase
due to the checkerboard pattern of vortices
and anti-vortices coexisting with the SC order. The chiral LRO is
demonstrated by the finite expectation value of chiralities %
\eqref{eq:chiral_op} as shown in Fig.~\ref{fig:mdfxy_low_T} (b) and Fig.~\ref%
{fig:mdfxy_mu_0.2} (b). The SC order is characterized by the quasi-LRO of $%
U(1)$ spins, where the spin-spin correlation function \eqref{eq:G} displays a
power-law decay as displayed in Fig.~\ref{fig:mdfxy_low_T} (d). The melting
of the vortex lattice undergoes two steps into the disordered phase with an
intermediate non-SC vortex-lattice phase. In the non-SC vortex-lattice
phase, the chiral LRO survives but the phase coherence between vortices is
destroyed. Such a two-step procedure has been extensively investigated in
the FFXY models~\cite{Villain1977,Villain1977_2,Song_2022}.

\begin{figure}[t]
\centering
\includegraphics[width=\linewidth]{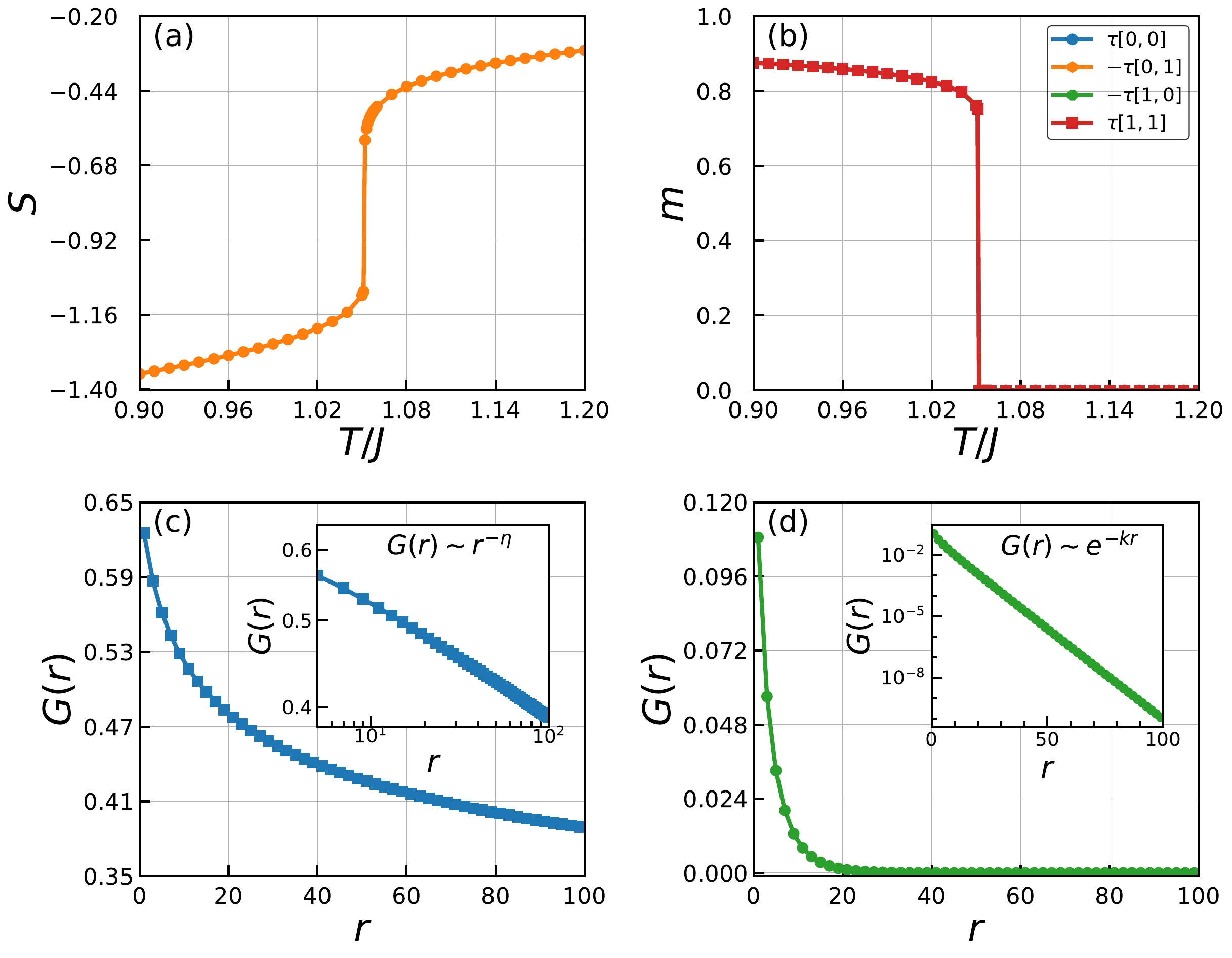}
\caption{ (a) The thermal entropy density as a function of temperature along
$\protect\mu=0.18J$. (b) The chirality on $2\times 2$ sublattice as a
function of temperature along $\protect\mu=0.18J$. (c) The spin-spin
correlation function at $T\simeq1.05J,$ and $\protect\mu \simeq0.18J$ in
vortex lattice phase shows a power-law decay. (d) The spin-spin correlation
function at $T\simeq1.06J$ and $\protect\mu\simeq0.18J$ in disorderd phase
displays an exponential decay.}
\label{fig:mdfxy_mu_0.2}
\end{figure}

Below the point $D$, the phase boundaries are determined by a combined
analysis of the entanglement entropy and free energy. We find that the
fixed-point equations have two different solutions across the critical point
depending on the initial states we start from. The proper solution is chosen
with a lower free energy density. As shown in Fig.~\ref{fig:mdfxy_EE} (c),
along the line $T=0.8J$, the entanglement entropy exhibits a peak at
$\mu\simeq 0.080J$ corresponding to the BKT transition and a discontinuous
jump at $\mu\simeq 0.153J$ associated to a first-order phase transition.
The free energy density of Fig.~\ref{fig:mdfxy_EE} (e) displays an inflection
point of a first-order transition at $\mu\simeq 0.153J$, demonstrating that the
entanglement entropy can serve as a powerful criterion for the determination
of the first-order phase transition. Besides, we find that the position of
the first-order transition is nearly unchanged with increasing bond
dimensions, in good agreement with the behavior of the entanglement entropy.
As the temperature decreases, the BKT transition line $CA$ and the
first-order transition line $CD$ become closer and finally merge into a
single first-order transition line $CB$ at the tricritical point $C$ with
$T\simeq 0.640J$ and $\mu\simeq 0.142J$. As shown in Fig.~\ref{fig:mdfxy_EE}
(d), along the line $T=0.64J$, the entanglement entropy shows a discontinuous
jump just above the peak position of $\mu\simeq0.142J$. The corresponding free
energy density is displayed in Fig.~\ref{fig:mdfxy_EE} (f) with an evident
cusp point.

Across the transition line $CD$, the vortex lattice melts directly into the
disordered phase via a first-order transition, where the chiral LRO and spin
quasi-LRO break down simultaneously. As is shown in Fig.~\ref%
{fig:mdfxy_mu_0.2} (a) and (b), both the thermal entropy density $S$ and the
chiral order parameter $m$ develop a discontinuous jump at the transition
point of $\mu\simeq 0.18$ and $T\simeq 1.052$. A comparison between the
spin-spin correlation functions across the line $CD$ is displayed in Fig.~%
\ref{fig:mdfxy_mu_0.2} (c) and (d). For a given temperature of $T=1.05J$ in
the vortex-lattice phase, the correlation function $G(r)$ displays a
power-law behavior. But in the disordered phase with $T=1.06J$, the
correlation function behaves in an exponential way. We should point out that
the existence of a novel continuous transition arising from the merging of
BKT and Ising transitions~\cite%
{Granato_1986,Granato_1987,lee_1991_2,li_1994,Nightingale_1995} is not found
here.

At low temperatures, the phase boundary $CB$ belongs to a first-order
transition between the vortex-lattice phase and the vortex-vacuum phase. As
shown in Fig.~\ref{fig:mdfxy_low_T}(b), when going down along $T=0.1J$ line,
the chiral order parameter $m$ exhibits a discontinuous jump to zero at
$\mu \simeq 0.128J$. Since the vortex fugacity is greatly suppressed by
decreasing the chemical potential $\mu$, the vortex density drops
dramatically, driving the system into the vortex-vacuum phase. Note that the
``vortex vacuum" just means that there is no excitation of free vortices but the
charge-neutral vortex-antivortex pairs can still be excited. The excitation of
vortex-antivortex pairs destroys the LRO of the $U(1)$ spins and gives rise to
the well-known BKT quasi-LRO state. As can be seen in
Fig.~\ref{fig:mdfxy_low_T} (c)-(d), the spin-spin correlation function displays
a power-law decay in both the vortex-lattice and vortex-vacuum phases. When
the temperature further decreases, the first-order transition line $CB$ behaves
in a linear way. Such a linear behavior is displayed in
Fig.~\ref{fig:mdfxy_low_T} (a), where the extrapolation to the zero temperature
gives $\mu=0.125J$ in the inset. The terminal point $B$ is determined at $T=0$
and $\mu=0.125J$, consistent with our previous analysis of the ground state.

Finally, the transition line $CA$ separating the vortex-vacuum and
disordered phase is the conventional BKT transition, driven by the
dissociation of vortex-antivortex pairs. The inverse process, when the
system is cooling from a disordered phase, pairs of vortex and anti-vortex
appear and further condensed into a square vortex lattice is analogous to
the theoretical proposal in ultracold Fermi gases \cite{Botelho_2006}.
The rich phase diagram of the modified XY model provides important insights
into the formation of the vortex lattice and the complex melting process. By
tuning the vortex chemical potential, the unconventional phase transitions
in SC lattice are investigated thoroughly in the orientational $U(1)$ phase
variables. A more comprehensive study should take into account the
positional order since the vortex lattice may also melt via the
Kosterlitz-Thouless-Halperin-Nelson-Young procedure~\cite%
{Halperin_1978,Nelson_1979,Young_1979,Zhang_SC_1993,Gabay_1993,Botelho_2006}.

\section{Discussion and outlook}

In this paper, we have developed a generic tensor network approach to study
the frustrated classical spin models with both discrete and continuous
degrees of freedom on a wide range of 2D lattices. The key point for a
contractible tensor network representation of the partition function is that
the emergent degrees of freedom induced by frustrations should be encoded in
the local tensors comprising the infinite network. In this way, the massive
degeneracy can be described by the interactions between emergent dual
variables representing a cluster of interacting spins under the constraint
of frustrations. We showed that a common process can be applied to the
construction of the tensor network based on ideas of emergent degrees of
freedom and duality transformations. We demonstrated the power of our method
by applying it to a large array of classical frustrated Ising models and
fully frustrated XY spin models on the kagome, triangular and square
lattices in the whole temperature range. Our tensor network approach turned
out to be a natural generalization of the previous solutions of frustrated
spin systems\cite{Vanhecke2021,Colbois_2022,Song_2022,Song_2023_2} but from
a more fundamental basis. Then the partition function is expressed in terms
of a product of 1D transfer matrix operator, whose eigen equation was solved
by the algorithms based on matrix product states rigorously. The singularity
of the entanglement entropy for the 1D quantum analog provides a stringent
criterion to determine various phase transitions with high accuracy. Apart
from the good agreement with previous findings, our numerical results offer
new clarification of the phase structure of the AF triangular XY model and
the modified XY model.

The generic tensor network approach provides a promising way to deal with
some remaining open questions on frustrated systems. First, our method
should be applicable to frustrated spin models with longer-range
interactions where emergent degrees of freedom play an important role in
characterizing the collective behavior. For example, a range of novel
classical spin liquid phase in the $J_1$-$J_2$-$J_3$ Ising model at the
fine-tune point can be understood by topological charges with the nearest
neighbor interaction and hence can be solved directly from our tensor
network approach\cite{Mizoguchi_2017,Tokushuku_2019,Tokushuku_2020}. Second,
the long-standing problems in uniform frustrated XY spin models may be
solved by our generic construction. All the frustration ratio $f\in [0,1]$
can be represented by a suitable gauge field on the lattice bond, which can
be further represented using the standard procedure. Finally, we should
point out that our construction should be extended to other models in any
dimension with emergent degrees of freedom. For instance, the classical
Heisenberg antiferromagnet\cite{Chalker_1992,Pitts_2022} may be investigated
in the future where the basis for the dual transformation should be
spherical harmonic functions. We believe that further development of the
tensor network approach of our work should lead to the solution of a number
of problems in frustrated systems that were difficult to solve previously.

\begin{acknowledgments}
The authors are very grateful to Tao Xiang for his stimulating discussions.
The research is supported by the National Key Research and Development
Program of MOST of China (2017YFA0302902).
\end{acknowledgments}

\appendix*

\section{Tensor network calculations of the physical quantities}

\subsection{Linear transfer matrix method}

Once the proper tensor network representations for the frustrated models are
obtained, the contraction of the infinite tensor network can be performed
efficiently. One of the best practices to contract a translation-invariant
tensor network in the thermodynamic limit is the algorithm of uniform matrix
product states where the leading eigenvector of the row-to-row transfer
matrix is calculated using a set of optimized eigensolvers\cite%
{Stauber_2018, Fishman_2018, Vanderstraeten_2019}.

Due to the emergent phenomena in the frustrated systems, the lattice
symmetry is usually spontaneously broken with a larger translation-invariant
unit composed of new degrees of freedom. The relevant 2D tensor network
should consist of a larger unit cell of multiple tensors that matches the
transitional symmetry. For example, a $2\times2$ plaquette structure of $O$
tensors is necessary to represent the checkerboard ground state of the FFXY
model on square lattices and a $3\times3$ structure for the triangular AF XY
model.

The fixed-point equation for the enlarged transfer operator can be
accurately solved by the multiple lattice-site VUMPS algorithm with only a
linear growth in computational cost\cite{Nietner_2020}. For a
transition-invariant cluster consisting of $n_x\times n_y$ local tensors,
the whole transfer matrix is formed by $y$ rows of linear transfer matrices
\begin{equation}
\mathcal{T}=T^{(y+n_y-1)}\cdots T^{(y)},
\end{equation}
where each row of the component transfer matrix is defined by
\begin{equation}
T^{(y)}=\mathrm{tTr}\left(\cdots O^{{(x,y)}} O^{{(x+1,y)}} \cdots\right)
\end{equation}
with $x=0,\cdots,n_x-1$, and $y=0,\cdots,n_y-1$. The transfer operator $%
\mathcal{T}$ can be regarded as the matrix product operator (MPO) for the 1D
quantum spin chain, whose logarithmic form can be mapped to a 1D quantum
system with complicated spin-spin interactions
\begin{equation}
\hat{H}_{1D}=-\frac{1}{\beta}\ln\mathcal{T}.
\end{equation}
In this way, the correspondence between the finite temperature 2D
statistical model and the 1D quantum model at zero temperature is
established.

\begin{figure}[tbp]
\centering
\includegraphics[width=\linewidth]{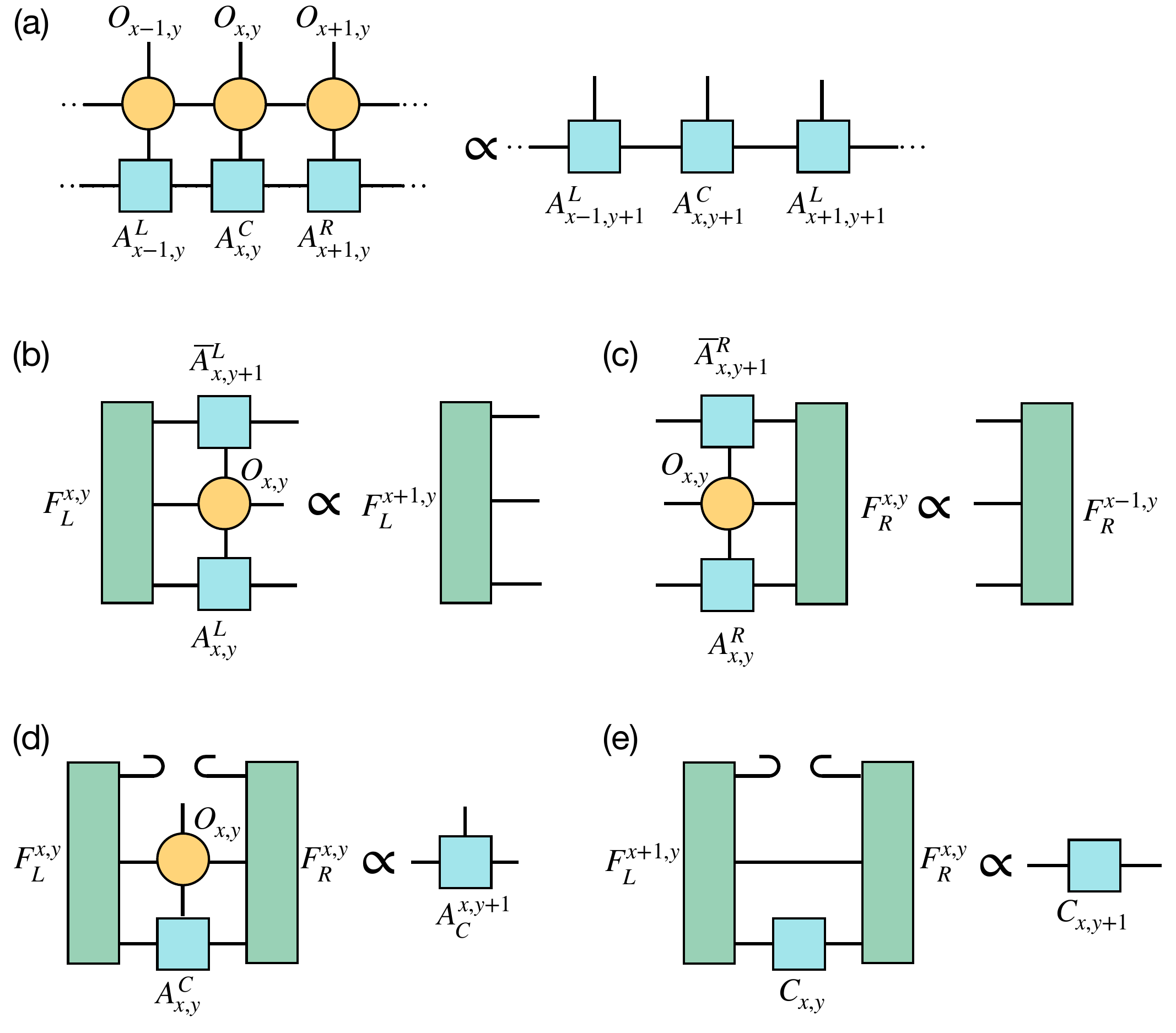}
\caption{ The key steps of the multi-site VUMPS algorithm. (b) and (c)
Eigen-equations to update the left and right environmental fixed points of
the channel operators. (e) and (f) Eigen-equations to update the central
tensors based on the new environment.}
\label{fig:vumps}
\end{figure}

The eigenequation can be expressed as
\begin{equation}
\mathcal{T}|\Psi(A)\rangle^{(y)}=\Lambda_{\max}|\Psi(A)\rangle^{(y)},
\end{equation}
where $|\Psi(A)\rangle^{(y)}$ is the leading eigenvector represented by
matrix product states (MPS) made up of a $n_x$-site unit cell of local A
tensors with auxiliary bond dimension $D$
\begin{multline}
|\Psi(A)\rangle^{(y)}=\sum_{x}\mathrm{Tr}(\cdots A^{i_{(x,y)}}
A^{i_{(x+1,y)}} \cdots) |\cdots i_{(x,y)} \cdots\rangle
\end{multline}
satisfying $A^{(x,y)}=A^{(x,y+n_y)}=A^{(x+n_x,y)}$\cite{Stauber_2018}. The
big eigenequation can be further decomposed into a set of smaller
eigen-equations displayed in Fig.~\ref{fig:vumps} (a) as
\begin{equation}
T^{(y)}|\Psi(A)\rangle^{(y)}=\Lambda_y|\Psi(A)\rangle^{(y+1)},
\end{equation}
with a total eigenvalue
\begin{equation}
\Lambda_{\max}=\prod_{y=0}^{n_y-1} \Lambda_y.
\end{equation}

The key process of the algorithm is summarized in Figs.~\ref{fig:vumps}
(b)-(e), including sequentially solving the left and right fixed points of
the channel operators
\begin{align}
\mathbb{T}_L^{(x,y)} F_L^{(x,y)} &= \lambda_{(x,y)} F_L^{(x+1,y)}, \\
\mathbb{T}_R^{(x,y)} F_R^{(x,y)} &= \lambda_{(x,y)} F_R^{(x-1,y)},
\end{align}
and the updating of the central tensors
\begin{align}
H_{A_C}^{(x,y)} A_C^{(x,y)} &= \lambda A_C^{(x,y+1)}, \\
H_C^{(x,y)} C^{(x,y)} &= C^{(x,y+1)}.
\end{align}
Note that, when solving the fixed point eigen equation (A.8)-(A.11), one may
not directly use the linear transfer matrix composed by the uniform local
tensor ${O}$, but the interior structure should be explored. This will
significantly reduce the computational complexity.

\subsection{Physical quantities}

\begin{figure}[tbp]
\centering
\includegraphics[width=\linewidth]{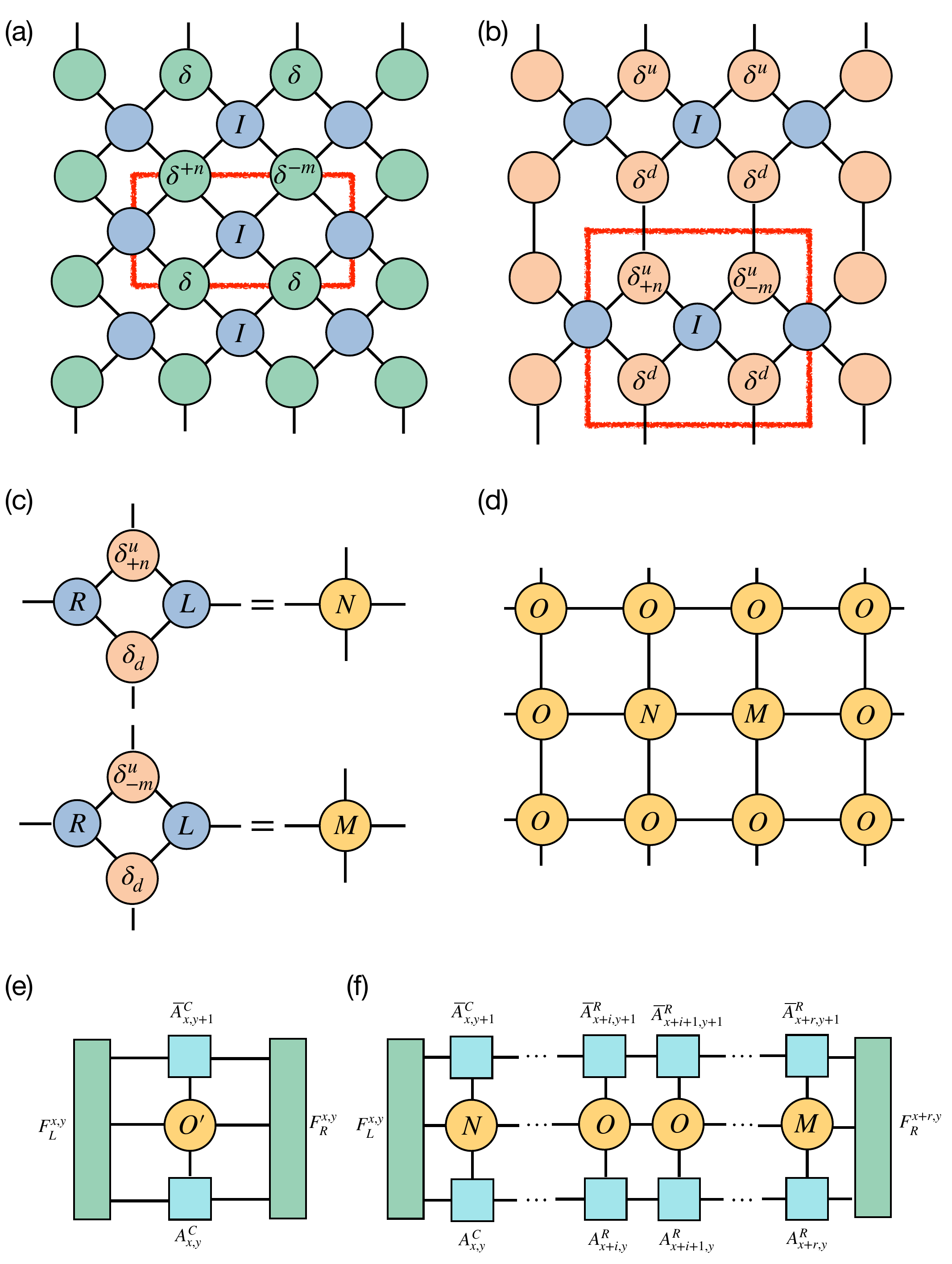}
\caption{ (a) The imbalanced delta tensors as a result of imbalanced
currents introduced by the local observables. (b) The vertical split of the
imbalanced delta tensors. (c) The construction of the impurity tensors from
imbalanced delta tensors. (d) Two impurity tensors are introduced into the
original tensor network. (e) Expectation of a local observable by
contracting the leading eigenvectors of the channel operators. (f) Two-point
correlation functions calculated by contracting a sequence of channel
operators. }
\label{fig:impurity_tensor}
\end{figure}

From the fixed-point MPS for the 1D quantum transfer operator, various
physical quantities can be estimated accurately. The entanglement properties
can be detected via the Schmidt decomposition of $|\Psi(A)\rangle^{(y)}$
which bipartites the relevant 1D quantum state of the MPO, and the
entanglement entropy can be determined directly from the singular values $%
s_\alpha$ as
\begin{equation}
S_E=-\sum_{\alpha=1}^{D}s_{\alpha}^2\ln s_{\alpha}^2,
\end{equation}
in correspondence to the quantum entanglement measure.

Moreover, the expectation value of a local observable can be evaluated by
inserting the corresponding impurity tensor into the original tensor network
for the partition function. The impurity tensors can be obtained simply by
introducing an unbalanced delta tensor to replace the original delta tensor
characterizing the constraints of sharing spins.

For Ising spins, the expectation value of a local spin at site $j$ can be
expressed as
\begin{equation}
\langle s_j\rangle=\frac{1}{Z}\sum_{\{s_i=\pm1\}}\mathrm{e}^{-\beta
E(\{s_i\})}s_j
\end{equation}
where $E(\{s_i\})$ is the energy of a state under a given spin configuration
$\{s_i\}$. The $s_j$ term just changes the Kronecker delta tensor from the
form of \eqref{eq:kroneck_delta} to
\begin{equation}
\delta_{s_1,s_2,\cdots,s_n}=
\begin{cases}
s_1, & s_1=s_2=\cdots=s_n \\
0, & \text{otherwise}%
\end{cases}%
.
\end{equation}

For XY spins, the expectation value of $\mathrm{e}^{iq\theta}$ can be
calculated by introducing imbalanced currents into the original delta
tensors from the conservation form of \eqref{eq:conserv_delta} to
\begin{equation}
\delta^q=\delta_{n_1+n_2+n_3+n_4+q,0}
\end{equation}
as displayed in Fig.~\ref{fig:impurity_tensor} (a). Accordingly, the
vertical splitting of the delta tensor in \eqref{eq:vsp_delta} should be be
modified to
\begin{equation}
\delta_{n_1+n_2+n_3+n_4+q,0}=\sum_{n_5}\delta_{n_1+n_2-n_5,0}^u%
\delta_{n_1+n_2+n_5+q,0}^d
\end{equation}
as shown in Fig.~\ref{fig:impurity_tensor} (b). Then the impurity tensors
can be constructed in the same way by including the imbalanced delta tensors
as depicted in Fig.~\ref{fig:impurity_tensor} (c). The tensor network
containing two impurity tensors is displayed in Fig.~\ref%
{fig:impurity_tensor} (d) as an example.

Using the MPS fixed point, the contraction of the tensor network containing
the impurity tensor is reduced to a trace of an infinite sequence of channel
operators, which can be further squeezed into a contraction of a small
network. As shown in Fig.~\ref{fig:impurity_tensor} (e), the evaluation of a
single variable is expressed as a contraction of only five tensors. And the
expectation value of the two-point correlation function
\begin{equation}
G(r)=\langle \cos(n\theta_i-m\theta_{i+r})\rangle
\end{equation}
can be reduced to a trace of a row of channel operators containing two
impurity tensors as shown in Fig.~\ref{fig:impurity_tensor} (f).

\bibliography{ref}

\end{document}